Enforcing Regulation under Illicit Adaptation

By

Andres Gonzalez Lira   and Ahmed Mushfiq Mobarak

August 2018

COWLES FOUNDATION DISCUSSION PAPER NO. 2143

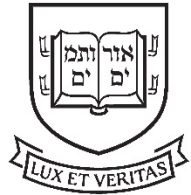

# Enforcing Regulation under Illicit Adaptation

Andres Gonzalez Lira[*]   Ahmed Mushfiq Mobarak[†]

August 11, 2018

**Abstract**

Attempts to curb illegal activity by enforcing regulations gets complicated when agents react to the new regulatory regime in unanticipated ways to circumvent enforcement. We present a research strategy that uncovers such reactions, and permits program evaluation net of such adaptive behaviors. Our interventions were designed to reduce over-fishing of the critically endangered Pacific hake by either (a) monitoring and penalizing vendors that sell illegal fish or (b) discouraging consumers from purchasing using an information campaign. Vendors attempt to circumvent the ban through hidden sales and other means, which we track using mystery shoppers. Instituting random monitoring visits are much more effective in reducing *true* hake availability by limiting such cheating, compared to visits that occur on a predictable schedule. Monitoring at higher frequency (designed to limit temporal displacement of illegal sales) backfires, because targeted agents learn faster, and cheat more effectively. Sophisticated policy design is therefore crucial for determining the sustained, longer-term effects of enforcement. Data collected from fishermen, vendors, and consumers allow us to document the upstream, downstream, spillover, and equilibrium effects of enforcement on the entire supply chain. The consumer information campaign generates two-thirds of the gains compared to random monitoring, but is simpler for the government to implement and almost as cost-effective.

[*]UC Berkeley, andres.gonzalez@berkeley.edu

[†]Yale University and NBER, ahmed.mobarak@yale.edu

[‡]We thank the National Marine Resource Authority of Chile, *Sernapesca*, for their cooperation in implementing a randomized controlled trial; *JPAL-Latin America* for providing research and data collection support; the Macmillan Center at Yale, the Center for Business and Environment at Yale, and *Sernapesca* for financial support; Gunjan Amarnini, Olivia Bordeu, Pascuala Domínguez, Angélica Eguiguren, Matthew Sant-Miller and Diego Verdugo for research assistance; Anjali Adukia, Ernesto Dal Bo, Ruben Durante, Ed Glaeser, Rick Hornbeck, Kyle Meng, Reed Walker, and seminar and conference participants at AEA 2018 Meetings, Berkeley-Haas, INSEAD, PUC-Chile, Sciences Po, University of Hawaii, UC-Berkeley Econ, Vienna Univ. of Economics and Business, Yale University, and North-east Environmental Economics Workshop for comments. This study was registered in the AEA RCT Registry: AEARCTR-0000822.

# 1  Introduction

Correcting market failures and improving economic efficiency often requires curbing undesirable behaviors of market agents who act to maximize their private benefits. Examples span actions that affect the natural environment, such as deforestation (Jayachandran et al., 2017), pollution (Duflo et al., 2013, 2018), or over-exploitation of natural resources (Stavins, 2011), actions that affect community health such as drunk driving (Banerjee et al., 2017) and open defecation (Guiteras et al., 2015), or actions that undermine government performance such as corruption (Fisman and Wei, 2004; Olken, 2007) and tax evasion (Carrillo et al., 2017; Fan et al., 2018). Enacting and enforcing regulations is the most direct strategy to deter such behaviors. Implementing this strategy requires strong institutions to enforce laws, plus sophisticated policing to track agents' reactions to enforcement, so that rules are robust enough to curb the undesirable behavior even when regulated agents try to 'game' the new system.

Effective enforcement is challenging precisely because the targeted agents will react to the new set of rules, finding loopholes that allow them to continue maximizing private benefits at the expense of others.[1] In many instances, it is therefore insufficient to evaluate the effectiveness of enforcement activities (such as an anti-corruption campaign) based on their immediate, short-run effects, before regulated agents have had a chance to adjust to the new regime. A more sophisticated evaluation will need to track the targeted agents' reactions, and the (sometimes unanticipated) strategies that they may deploy to circumvent the regulation. This will also typically require innovations in measurement of the illegal activity, and clever data collection that does not rely on the regulators' administrative records.

We carry out such an investigation on the effects of – and the limits of – enforcement in the context of illegal sales of the critically endangered Pacific hake fish (*merluza*) in urban markets in Chile. The government of Chile has instituted a ban on fishing and sales of hake during September each year, when the fish reproduces. Catching hake during that period is especially ecologically destructive. We implement a randomized controlled trial (RCT) in which government agents monitor and penalize vendors that sell illegal fish, while we surreptitiously monitor vendors' reactions to that enforcement by deploying "mystery shoppers" to search for illegal hake in markets. We also randomly varied the predictability and frequency of enforcement activities to study whether certain innovations in enforcement design are more or less effective in curbing the undesired behavior, accounting for the targeted agents' reactions to the new treatment regime. We collect data along the entire supply chain: surveying fish vendors, consumers who buy from those vendors, and fishermen[2] who sell to these vendors, to study the upstream and downstream effects of this intervention.

---

[1]For example, Carrillo, Pomeranz, and Singhal (2017) show that when the Ecuadorian tax authority improves the quality of their information on firms' tax revenues, the firms react by raising their estimates of costs in line with the revised revenue estimates, to keep total tax payments unchanged. Blattman et al. (2017) shows that intensive policing pushes crime around the corner, with null impacts on overall violent crimes.

Finally, we implement a consumer information campaign which allows us to benchmark the effects of enforcement against this demand-side strategy that is easier for policymakers to implement. When it is difficult or expensive to enforce rules, less direct strategies such as information campaigns designed to change social norms around the undesirable behavior, or consumer-marketing that appeals to regulated agents' sense of fairness, or encouraging third-party reporting may be more reliable or more cost-effective. For example, Chetty et al. (2014) partners with the Bangladesh tax authorities in an attempt to change social norms to encourage firms to pay taxes (as opposed to enforcing tax laws directly), and Guiteras et al. (2015) attempt to change social norms around toilet use (as opposed to directly banning the dangerous practice of open defecation). The information campaign in the present study was designed to educate consumers about the environmental risk associated with over-fishing of hake and to discourage the consumption of illegal hake during the September ban period. The demand-side and enforcement interventions may even be complementary: If vendors react to the enforcement by hiding their illegal hake sales, then informed consumers may be an important second line of defense. Our 2x2 experimental design with enforcement, information, both and neither can test for such complementarities.

Since we are tracking an illegal activity, our main outcome variables for the evaluation do not rely on official statistics but are collected through a "mystery shopper" methodology. We sent trained surveyors who look like typical shoppers to each market to pose as buyers and (try to) purchase fish during the ban. We collected data on whether it was possible to buy hake and its market price, as well as the price and availability of substitutes to hake fish. Vendors would have an incentive to under-report illegal hake sales to enumerators, which is why the mystery shopper approach improves the credibility of our evaluation data.

We also conducted consumer surveys before and after the interventions, to gather data on changes in fish consumption behavior and consumer knowledge about the ban. We repeated the mystery shopper and consumer surveys six and nine months after the interventions to gauge long term effects. We mapped all spatial and market relationships between vendors and fishermen to study spill-overs across markets. Finally, we surveyed the fishermen who supply to these markets to explore whether the enforcement and information interventions implemented "downstream" (at the point of sale from vendors to consumers) traveled "upstream" the supply chain of fish. It is ultimately the fishermen who make the ecologically sensitive decisions in the seas. These multiple sources of data allows us to provide a more comprehensive evaluation of the full range of effects up and down the supply chain. Our evaluation was conducted at scale with the Chilean government, covering 70 districts that cover all major markets in Central Chile where the majority of hake is sold. We can therefore report on equilibrium outcomes such as prices of hake and its substitutes. Finally, we map spatial, social and supply chain relationships between markets and fishermen in order to study spillovers across markets.

---

[2] We use the gendered term "fishermen" because every single fisher we interviewed in Chile was a man.

We find that many vendors continue selling illegal hake during the ban, but both the enforcement treatment and the information campaign reduced the sale of illegal hake fish. Declines in hake sales in treatment areas during the ban period are twice as large as the decline in the control markets. Enforcement generates slightly larger reductions in hake sales compared to the information campaign. Our mystery shopper data provides clear evidence that vendors react to enforcement activity by engaging in new practices designed to circumvent our attempts to levy penalties. Many vendors do not display the hake openly during the ban, but are willing to sell our mystery shoppers hake fish that is hidden from plain view. They also start keeping the hake on ice and claiming that the fish on display was caught in August when it was still legal to do so. These reactions attenuate the effects of enforcement on the *true* availability of illegal hake in markets.

Random variation in specific design features of the enforcement strategy reveals that government agents monitoring vendors on a predictable schedule is not effective. Vendors find it easier to cheat (through hiding and freezing) when the enforcement visits become predictable. We also tried increasing monitoring frequency and intensity to better contain temporal and spatial spillovers to other days of the week or other nearby markets. But this strategy backfired because it evidently allowed fish vendors to learn the monitoring routines more quickly and react with greater hiding and freezing of illegal fish. In the subset of locations where we send monitors on an unpredictable and less frequent schedules, vendors were not able to learn and adjust as quickly, and this resulted in large reductions in hake sales, even accounting for the hiding and freezing response.

Our main contribution is to provide empirical evidence on the real world challenges to imple-menting policies that are designed to address negative externalities and collective action failures. Our focus on policy implementation is related to a literature on the theory of regulation that confronts the possibility of subversive behavior by agents (Glaeser and Shleifer, 2003), and theoretical work in public finance on optimal (and second best) policy design to manage public goods (Burgess and Stern, 1993). Our empirical work complements this literature by grappling with the complexities of real world policy implementation.

Becker (1968) seeded a literature on how regulated agents respond to changes in their perceived probability of detection (Townsend, 1979; Harrington, 1988). Okat (2016) and Eeckhout et al. (2010) make theoretical points that are closely related to our empirical results. For example, Okat (2016) predicts that unpredictable and less frequent enforcement hinders or delays agents' learning about the weaknesses of the auditing process: *"individuals audited potentially learn how to exploit the weaknesses inherent in any audit methodology if they face the same method many times"*.

Our paper is related to the empirical literature on the effects of monitoring and penalties (Shimshack and Ward, 2005; Gray and Shimshack, 2011; Hansen, 2015; Pomeranz, 2015). Like our consumer information campaign, many other papers have evaluated indirect strategies in pursuit of social goals, in environments where enforcement is expensive or difficult.[3]

The paper is organized as follows. In section 2 we describe the background and experimental design. In section 3 we introduce the theoretical framework and testable predictions. Section 4 describes the different data sources and present summary statistics. Section 5 presents the empirical strategy and results, section documents spillovers and market equilibrium effects, and section 8 concludes the paper.

## 2 Background and Experimental Design

### 2.1 Context

With around 4,000 miles of coastline, Chile is one of the top ten fish producers in the world (FAO, 2014). However, as in many other low and middle-income countries, the marine ecosystems have been threatened by over-fishing. The Chilean government has passed various regulations to protect

---

[3] Jin and Leslie (2003), Reinikka and Svensson (2005), Alm et al. (2009), Shimeles et al. (2017) and Kollmuss and Agyeman (2002)

threatened species over the last 20 years, including restrictive fishing quotas and fishing ban periods. However, the fish population has continued to decrease, with 72% of species rated as overexploited or *collapsed* by 2015 (Subpesca, 2015). These regulations have been difficult to implement and enforce.

The majority of people carrying out fishing activities are small-scale and artisanal fishermen, which makes the sector very difficult to regulate. Small-scale fishermen contribute almost 40% of the national fishing volume, and up to 75% of the hake fish market. The 30% of fish produced by industrial fisheries and the remaining 30% from aquaculture are much easier to regulate. Artisanal fishermen are organized in fishing villages called *Caletas*. Around 76% of the caletas are located in rural areas along the extended Pacific coast, and they are highly spatially dispersed (Subpesca, 2013). Their geographic dispersion, informality, and the small-scale of operations of each individual fisherman make it difficult for the government to monitor their activities. The absence of alternative income-generating activities for these fishermen has also make it difficult to change the norms regarding "appropriate behavior" in this industry. Furthermore, poor small-scale fishermen do not readily accept government-imposed restrictions, and they have organized and unionized to create political opposition to government policies that restrict fishing. The number of small-scale fishermen has actually increased over the last decade (Subpesca, 2013), which suggests that the regulatory efforts have had a limited effect on this group.

The Pacific Hake is the fish low and middle-income Chileans consume most, and also one of the most important sources of protein for this population. Hake is now critically threatened due to illegal over-fishing. The domestic hake market is served entirely by the domestic supply. Imports and exports of hake are quite uncommon. In an effort to protect the hake population, the Chilean National Marine Authority (*Sernapesca*) and the central government have enacted various policies including restrictive fishing quotas and a one-month ban on fishing and selling hake during the fish's September reproduction cycle. Due to difficulties in enforcing the ban, the hake population has continued to shrink, and the government now estimates that the population is 18% of its long-term sustainable level (Subpesca, 2015).

## 2.2 Supply Chain of Illegal Fish

### 2.2.1 *Caletas:* Coastal Villages where Artisanal Fishermen Bring in their Catch

Most of the illegal hake fish is captured by small-scale rural fishermen operating out of hundreds of *caletas*. Each *caleta* contains between 10 and 100 fishing boats. Boats are about 20 and 30 feet in length, and operated by two to three fishermen (see Figure A.1). The fishermen operating out of a *caleta* are organized as a union to internally distribute the fishing rights allocated to that *caleta*. In practice, each fisherman captures illegal, undeclared fish beyond the allocated quota. WWF (2017) estimates that the amount of hake fished by small-scale artisanal fishermen are between 3.8 and 4.5 times the legal quota. As a result, the artisanal sector is responsible for 75% of the hake fish supplied in the market, even though they hold only 40% of the "official" hake quotas. Our experiment focuses on fish harvested by this sector.

Fishermen go fishing using artisanal boats and nets at night and sell fish after sunrise. They are able to target specific fish types by varying the location and depth at which the nets are dropped. The fish is sold directly at the docks to three types of buyers: (1) fish vendors who buy the fish to sell them in local markets, (2) intermediaries who supply fish to vendors located in places further from the coast, and (3) households who live close to the *caleta* and buy the fish for their own consumption. There is very little use of ice and refrigeration at this point in the supply chain. The fish that vendors sell in local markets is typically fresh, and captured the night before. Table C.3 in the Appendix describes *caleta* characteristics.

### 2.2.2 *Ferias:* Outdoor Markets where Hake is Sold

The majority of hake-fish sales to final consumers, especially low- and middle-income ones, occur in *ferias*, which are outdoor markets organized by municipalities. Each vendor pays a fee every six months to rent a selling spot in the market. In addition to fish, *ferias* sometimes contain stalls offering fruit and vegetables, clothes and other products.

Ferias are typically navigable only by foot, and each feria therefore serves a limited geographic area of surrounding neighborhoods. To cover more neighborhoods, the vendors rotate between different *ferias* in a pre-set pattern - typically setting up in the same location twice a week. For example, they may sell at a first feria every Sunday and Wednesday, at a second feria every Tuesday and Friday, and at a third feria every Thursday and Saturday. The group of vendors who move together across neighborhoods is called a *circuit*. The semi-annual fee paid to the municipality covers the vendor's inclusion in the entire circuit, so the same group of vendors typically rotate across neighborhoods all together. Vendors are not allowed to sell in public places other than *ferias*.

Each municipality typically organizes one circuit of vendors. Large municipalities may have more than one circuit. In such cases, the municipality area is divided in such a way that there is no geographic overlap between circuits. Figures A.2 and A.3 in the Appendix provide visual examples of ferias and circuits. Table C.1 describes observable characteristics of ferias' fish stalls.

## 2.3 Experimental Design

This study was implemented in close collaboration with the Chilean National Fish Service (Sernapesca), who has the ultimate regulatory authority over fishing activities in the country. Our implementing partner's goal from this project was to limit hake fishing, sales and consumption during the September ban. It is practically and politically very difficult for them to directly regulate fishermen, because their activities occur out in the water at night, and because the fishermen operating out of the geographically dispersed *caletas* are politically organized. *Sernapesca* therefore expressed an interest in exploring options to better regulate the fish sales at *ferias* where hake is most commonly sold.

### 2.3.1 Sample

We conduct our experiment in the five central regions of Chile, which is home to 74% of the Chilean population. The caletas located along the coastal villages and cities scattered across these five regions account for 98% of all hake fish harvested in Chile. We conduct our experiment in all ferias in these regions except for the city of Santiago.[4]

An important benefit of conducting the experiments at such a large and comprehensive scale is that it allows us to track any displacement of illegal hake sales towards control markets, because all potential markets (including ones where the interventions were not applied) are in our database.

---

[4]Santiago is unique in that there is one big centralized fish market called *Terminal Pesquero Metropolitano* (TPM) where vendors buy from intermediaries to re-sell at neighborhood ferias. TPM is already well-monitored by *Sernapesca*, and our interventions therefore did not need to be implemented there.

This allows us to trace the market-level equilibrium effects of our interventions. We collected data on the universe of circuits in our sample area, and from every fish vendor operating in those circuits. We mapped all ferias served by each caleta where the fish are caught. The unique long and thin geographic shape of Chile means that ferias are generally located very close to the caletas from where they source fish (22 miles away on average). This made it relatively easy to connect vendors to the fishermen they source from, and trace how the effects of our interventions are transmitted along the supply chain for hake fish.

There are 280 ferias (fish markets) operating in the 70 municipalities in our sample, and these ferias are organized into 106 separate *circuits*. In order to identify and map all existing ferias and circuits, we combined administrative data from multiple sources (Ministry of Economics and Sernapesca) along with information gathered from phone conversations with staff in every municipality. We then used Google Maps to define the consumer "catchment area" for each feria. We identify the neighborhoods which are likely served by each feria, considering the walking distance and road accessibility from the neighborhood to the feria, as well as the residential versus commercial/industrial characteristics of the neighborhoods. The location of the ferias and their organization as circuits were important for the design of our enforcement intervention. The definitions of the residential neighborhoods and their connections to each feria were important for the design of our consumer information campaign.

### 2.3.2 Interventions

This study experimentally evaluates the effects of two complementary interventions that aimed to reduce illegal sales of hake during the September ban period. These interventions were designed to affect:

1. The **supply** of hake by monitoring vendors and enforcing penalties on those found to be selling illegal hake.

2. The **demand** for hake through an information campaign designed to sensitize consumers about this environmental problem, and discourage hake consumption during the ban.

### 2.3.3 Design of Enforcement Intervention

The supply-side enforcement intervention deployed government monitors from *Sernapesca* to periodically visit ferias where fresh hake is usually sold, and levy fines if vendors are caught illegally selling hake during the September 2015 ban period. The punishment for illegal sales is a US $200 fine plus confiscation of the illegal fish. Sernapesca should ordinarily conduct some of these monitoring visits anyway, but as a part of this randomized controlled trial, they agreed to conduct more intensive monitoring consistently, at specific locations and according to schedules defined by the research team. They also shared detailed information on their monitoring activities and the identities of penalized fish vendors with the research team.

We anticipated that fish vendors would react to the enforcement activity by devising new defensive strategies that would help them avoid paying fines while continuing to sell hake in September. We introduced random variations in the enforcement policy design to investigate whether specific design variations make enforcement more or less effective in the presence of agents' efforts to circumvent the policy:

1. *Predictability:* We randomly varied the ease of predictability of the enforcement. In some areas, Sernapesca monitors followed a consistent schedule (e.g. M,W at 9am) that became easily predictable, while in other areas, they were asked to follow a less predictable schedule defined by the research team. The latter is a more expensive enforcement strategy because it requires having monitors on-call for longer windows. This strategy was practically more difficult for Sernapesca to implement.

2. *Intensity:* We randomly varied the intensity of the enforcement treatment at the circuit level, so that some groups of vendors only received one visit per week, while others were visited multiple times at the various locations in the city where they set up on different days of the week. Increasing the frequency of monitoring visits is more expensive, but our thinking was that it may limit vendors' ability to relocate illegal hake sales spatially and inter-temporally during the week. On the other hand, it may also accelerate vendors' "learning curve" about the nature of the 2015 enforcement, and devise effective defensive strategies more quickly.

Enforcement activities were randomized at the circuit-level, covering all 106 market-circuits. This randomization was stratified to ensure balance with respect to a few important spatial and market characteristics: (a) Whether the circuit was located in a coastal municipality, (b) Whether the circuit was the only one operating in its municipality, and (c) whether the circuit served geographically isolated communities.[5]

### 2.3.4 Design of Information Campaign

The demand-side intervention was a marketing campaign designed to inform consumers about the September ban on hake sales. Sernapesca distributed letters, flyers and hanging posters in the residential neighborhoods randomly assigned to this intervention. The letters, flyers and posters informed consumers about the decline in the hake population, and the ban on hake sales in September. Appendix B shows examples of flyers and the letter.

We identified whether each neighborhood was in the catchment area for each of the ferias in our sample, using our mapping exercise described in section 2.3.1. We used the location of major roads and crossings to define boundaries of neighborhoods, and attempted to divide the municipality up such that the population-size of neighborhoods would be roughly equal. We conducted this intervention in the 48 most populated municipalities, and identified 270 distinct neighborhoods in those municipalities. Figure B.3 provides example maps. The randomization procedure was as follows:

1. First, 18 of the 48 most populated municipalities were assigned to a high saturation information treatment, 17 to a low saturation information treatment, and the remaining 13 municipalities did not receive the letters, flyers or posters. "High saturation" was defined to be a case where two-thirds of the neighborhood in the feria's catchment area would receive the letters,flyers, and posters. In the low saturation treatment area, only one-third of the neighborhoods received those mailings. We randomly varied the proportion of neighborhoods receiving the treatment to examine whether there are larger changes in norms regarding the acceptability of inappropriate or socially harmful behavior when households observe that many of their neighbors simultaneously receive the same information about the illegality of hake consumption.

---

[5]We originally intended to apply this randomization to 153 circuits. However, 40 of those circuits did not have any fish-stalls. The mystery shoppers could not visit another 7 circuits for logistical reasons. These two sources of attrition are not correlated with any observable characteristics, nor with the treatment assignment.

2. Second, specific neighborhoods within each high or low saturation information treatment area were randomly chosen to receive the treatment.

3. Third, we randomly selected around 200 addresses in each of 102 neighborhoods, and mailed out letters to each of those 20,400 addresses. 200 letters cover roughly 15% of all potential addresses in a representative neighborhood. Based on information from the postal service, we subsequently learnt that at least 13,000 letters were correctly delivered.[6]

80,000 flyers were distributed by trained field personnel to people walking in the streets, and directly to households within the 102 treated neighborhoods. 3,000 posters were placed around treated neighborhoods where they would be publicly visible, such as at bus stations, community centers, and street intersections.

### 2.3.5 Cross-Randomized Experimental Design

The enforcement treatment and the information campaign were cross randomized in a 2x2 experimental design so that we could study potential complementarities between the two approaches. Table 1 lists the number of circuits assigned to each of the four treatment cells.

Table 1: Treatment Assignment

|  | No Enforcement | Enforcement | Total |
|---|---|---|---|
|  | N | N | N |
| No Information Campaign | 9 | 41 | 50 |
| Information Campaign | 14 | 42 | 56 |
| Total | 23 | 83 | 106 |

This table lists the number of circuits assigned to each experimental cell jointly defined by the Information Campaign (row) and the Enforcement treatment (column)

The majority of markets were assigned to Enforcement because that column contains additional sub-treatments in which we conduct experiments on variation in enforcement policy design. Those variations in predictability and frequency of enforcement visits were cross-randomized so that we have sufficient statistical power to study the effect of each variation, one at a time. Table 2 shows the number of circuits assigned to each sub-treatment cell. To study the effects of predictability of enforcement, we will compare the 39 circuits where Sernapesca monitored on a predictable schedule against the 44 circuits where they monitored on an unpredictable schedule. Similarly, to study the effects of enforcement intensity, we will compare the 34 circuits assigned to high-intensity against the 49 circuits assigned to low-intensity.[7]

---

[6]Although 13,000 were explicitly tracked, it is likely that around 16,500 were actually delivered, because the postal service did not receive any delivery failure notice in those cases. We inferred and constructed addresses using Google maps, and many of those addresses did not actually exist. That was a leading cause of delivery failure.

Table 2: Enforcement Sub-treatments

|  | High Intensity Enforcement N | Low Intensity Enforcement N | Total N |
| --- | --- | --- | --- |
| Predictable Enforcement Schedule | 19 | 20 | 39 |
| Unpredictable Enforcement Schedule | 15 | 29 | 44 |
| Total | 34 | 49 | 83 |

This table lists the number of circuits assigned to each experimental cell jointly defined by the row and column headers

Tests of the information campaign saturation effect (i.e. proportion of neighborhoods around markets that are simultaneously sent letters and flyers), will compare the 30 circuits randomly assigned to a low-saturation campaign (where a third of neighborhoods received letters and flyers), against the other 26 to a high-saturation campaign. We are able to control for other dimensions of random assignment whenever we focus on the effects of one particular dimension.

## 3  Theoretical Predictions

In theory, the enforcement treatment increases fish vendors' (perceived) probability of being detected and fined when selling illegal hake. The supply curve for hake in September should shift to the left, which will reduce the quantity of hake available in the market, and increase the market price. The subset of vendors who continue selling illegally will need to be compensated with higher prices for the increased risk they take.

The information campaign was designed to deter consumers from buying illegal hake. If successful, the demand curve should shift to the left, with fewer consumers purchasing at any given price. This should also decrease the quantity of hake sold in treated markets. Equilibrium prices should fall with this treatment. In summary, both the enforcement and information treatments should decrease the quantity of hake sold, but the net effect on price when both treatments are administered is ambiguous.

---
[7]The probability of assignment to low-intensity enforcement and to un-predictable schedules was a little higher compared to other cells. In our analysis, we will control for these differences.

## 3.1 Predictions on Fish Vendors' Reactions to our Interventions

We anticipated that vendors would react to the enhanced monitoring, or to more informed consumers, and would try to continue selling hake illegally by circumventing our enforcement and informational strategies. Our data collection and treatment variations were designed to shed light on those reactions as well:

1. Vendors may hide the illegal fish they sell in containers in the back, rather than display the fish openly. Another reaction we had not anticipated, but learnt about quickly from our "mystery shoppers", is that vendors try to pass off the hake as legal by keeping it on ice and claiming that it is frozen fish caught legally in August.

2. Vendors may also displace hake sales from monitored *ferias* to un-monitored *ferias* on a different week-day.

3. We will report results on vendors' propensity to sell fresh hake openly, sell hidden hake, or sell 'frozen' hake in response to our randomized treatments. What determines vendors' choices to sell visible or frozen or hidden hake? Icing hake to give it a frozen look reduces the quality/taste and appearance of freshness of the fish, which could lead to lower demand or lower prices. Hiding fish may depress demand because vendors have to rely on customers to explicitly ask for the fish. Displaying fish at their stall is the main advertising strategy that vendors use, so hiding is costly. On the other hand, selling visible fish puts the vendor at risk for a fine levied by *Sernapesca*, or complaints from environmentally-conscious consumers. Vendors' decisions on whether and how to sell hake will depend on the relative loss in profits due to these factors, and the tradeoff between reduced prices and demand from freezing or hiding, versus the expected cost of fines. The interventions may raise the expected cost of fines.

4. Vendors will need to learn about Sernapesca monitoring, before deciding whether to engage in defensive strategies like hiding and freezing. The speed and ease of learning about the nature of the enforcement will vary depending on how frequently Sernapesca officials visit, and whether those visits occur on a predictable schedule. It may also depend on whether consumers are informed, because hiding illegal activity from both clients and regulators may be much more complicated than hiding from regulators alone.

While monitoring on unpredictable schedules is more complicated and costly for the regulator,[8] it may prevent vendors from learning quickly about how to how displace illegal hake to a different day of the week to circumvent the enforcement efforts. We would expect the random monitoring visits to be more successful at reducing illegal hake sales than monitoring on a predictable schedule when vendors have the option to hide or freeze the hake, or displace sales to other market days.

Increasing the frequency of monitoring visits in our context implies that we visit the same vendors in multiple locations in their *circuit*, where they set up shop on different weekdays. We anticipated that this would make it more difficult for vendors to displace illegal sales from one week-day to another, from one of their market locations to another. However, high frequency visits can also allow vendors to learn about the regulators' schedule faster, and undertake effective defensive investments (Okat, 2016). They also create more vendor-regulator interactions, and allow vendors to learn quickly about the loopholes and weaknesses in the enforcement strategy, such as Sernapesca's unwillingness to penalize sales of hake that are on ice and plausibly caught in August. Increasing the frequency of enforcement therefore has an ambiguous effect on the sale of illegal hake if it is difficult for the regulator to detect and penalize behaviors like hidden or frozen hake sales.

Finally, while the demand-side information campaign and the supply side enforcement are normally thought of as substitute strategies to deter undesirable behaviors, the two approaches can become complementary in the presence of hidden and frozen sales. Hiding illegal hake only works well if consumers are willing to purchase hake in September. Putting fresh hake on ice can create plausible deniability when facing regulator fines, but may not be sufficient to fool consumers. The presence of an information campaign can therefore enhance the effectiveness of enforcement in this context.

---

[8]Administrative data from Sernapesca suggests that maintaining unpredictable schedules increase their program costs by 10% over and above a fixed schedule.

## 3.2 Predictions on the Nature of Spillover Effects

Enforcing the ban on hake sales can change the demand for other substitute species of fish. Vendor stalls often offer more than one type of fish, and hake is the only fish that is banned in September. If our interventions are successful, then consumers may choose to substitute to consuming other fish, and fishermen may substitute to catching those other fish. We might expect to see an increase in the quantity of substitute fish purchased in treatment areas. Prices of substitutes may also change if treatment and control markets are somewhat segmented. The direction of the price change will depend on the strength of the consumer response relative to the fishermen response.

Changes in consumer and vendor behavior may have an upstream effect on fishermen in *caletas* via the fish supply chain. If the quantity of hake fish sold in *ferias* decreases in September, that should affect the behavior of fishermen who serve those vendors. We will map the fish supply chains to empirically examine effects further up the supply chain.

Finally, control areas may experience some spillover effects. There are four possible channels of spillovers in our context:

1. Treated consumers who learn about the ban may inform other untreated consumers.

2. Treated vendors may share information with other vendors operating in control areas.

3. Some *caletas* supply fish to both treatment and control *ferias*. If fishermen switch their fishing strategy in response to treatment, supply of fish in control markets may also be affected.

4. If the information campaign is potent enough to decrease aggregate demand and the price of hake, there may be a general equilibrium spillover effect in control markets that enhances the direct effect of the treatment. On the other hand, if enhanced enforcement increases the price of fish, then the general equilibrium price spillover will undermine the direct effect of treatment.

It is important to understand the nature and direction of these spillovers, because the program effects might be under-estimated under certain types of spillovers. Or worse, if fishermen simply sell more illegal hake to control areas in response to treatment, then we will observe a treatment-control difference even if the true effect of the enforcement and information campaigns on aggregate hake fishing is nil. We will therefore collect data on spatial, supply chain, and personal relationships between treatment and control markets in order to analyze (and control for) these potential market spillovers.

# 4 Data

We conducted several different surveys to evaluate the effects of these interventions. We sent "mystery shoppers" to the fish markets over two different periods to surreptitiously gather information about hake availability, once during September 2015 (during the ban), and again six months later in March 2016. We conducted two rounds of surveys of consumers during those same two periods. We also surveyed fishermen at *caletas* and vendors at *ferias* to map the fish supply chain and investigate spillovers. Figure 1 describes the timing of the interventions and of data collection activities between August 2015 and August 2016. In total, seven different data sources are used in the analysis.

Figure 1: Timeline of Interventions and Data Collection

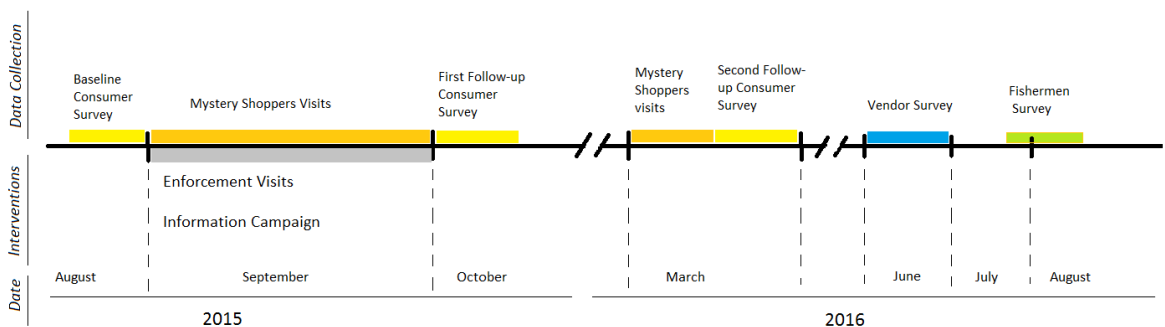

## 4.1 Mystery Shopper Surveys

We are interested in studying whether fish vendors engage in an illegal activity during the September ban. To obtain reliable measures of their behavior, the vendors cannot know that they are monitored, and this poses an interesting data collection challenge. To get around this problem, we deploy *mystery shoppers* to the fish markets, who look and act like typical customers, and who were trained to pose as buyers and (try to) purchase hake fish from the vendors. These trained surveyors gathered information on whether it was possible to buy hake, and on the market price of the fish. The mystery shoppers were also instructed to collect information on what else (e.g. other varieties of fish) was available for sale at the fish stalls and their prices, and to note down what was being purchased by other shoppers in their presence.[9]

The research team trained 29 mystery shoppers who circulated amongst all markets in our sample. Each circuit was visited, on average, three times during the September ban. To study longer-term effects of our interventions, we conducted an additional round of mystery shopper visits in March 2016, six months after the September ban period.

### 4.2 Consumer Surveys at Fish Markets

We also surveyed consumers before and after the ban period. A separate team of enumerators (distinct from our mystery shoppers) stopped consumers close the points of entry and exit for the fish market, and asked questions with a survey instrument in hand. To encourage unbiased responses, consumers were not asked to provide any personal identifiable information, and we only inquired about the list of food purchased in the feria in the past month - avoiding asking direct questions about the consumption of hake. We asked consumers to provide a sense of their home location on a map, so that we could match their residence to the neighborhoods assigned to the information treatment.

Around 4000 consumers were surveyed in October 2015, soon after the ban ended. Another 4000 surveys was completed in March 2016 to study longer-term changes in consumer behavior.

### 4.3 Survey of Vendors at Fish Markets

We surveyed fish vendors in every market in our sample in June 2016 (outside the ban period). We asked vendors about the suppliers and intermediaries they source their fish from, so that we could map out the supply chain. We also asked vendors about their contacts with fish vendors who operate in other circuits, in order to study spillover and network effects.

---

[9]Using this methodology, it was not possible to collect information about the total quantity of hake on offer or being sold, because that would be unnatural for a typical shopper to ask about, and it would have made the vendors suspicious. The main outcome variable that this survey therefore produces is an indicator for whether it was possible to buy hake from that vendor. The mystery shoppers also noted down general characteristics of the stall. They collected qualitative information about the behavior of the fish vendors, including conversations occurring at that moment. This is how we learnt about the practice of selling "frozen hake", where the vendor kept the fish on ice and claimed that it was caught legally in August. Many of those same vendors admitted to our mystery shoppers that the "frozen" fish was in fact, fresh.

## 4.4 Survey of Fishermen at Coastal Fishing Villages

To understand whether the effects of our interventions were transmitted upstream via the supply chain, we conducted a survey of fishermen in every coastal village in the region where hake fish is caught and distributed. This survey was carried out during July-August 2016, outside of the ban period. We surveyed 231 fishermen from 74 fishing villages (caletas). Figure A.4 in Appendix A.2 contains a map of all caletas and fish markets.

Surveying fishermen was valuable for two reasons. First, the interventions were designed to ultimately reduce illegal fishing, so understanding the activities of the fishermen is essential for public policy. Second, the treatment effects may have spilled over to control areas if treatment and control markets are served by the same fishing village. Understanding these supply-chain connections are important for analyzing spillovers. To clarify these connections, Figure 2 organizes our interventions and data collection activities along the supply chain for fish.

Figure 2: Interventions and Data Collection at different Points along the Fish Supply Chain

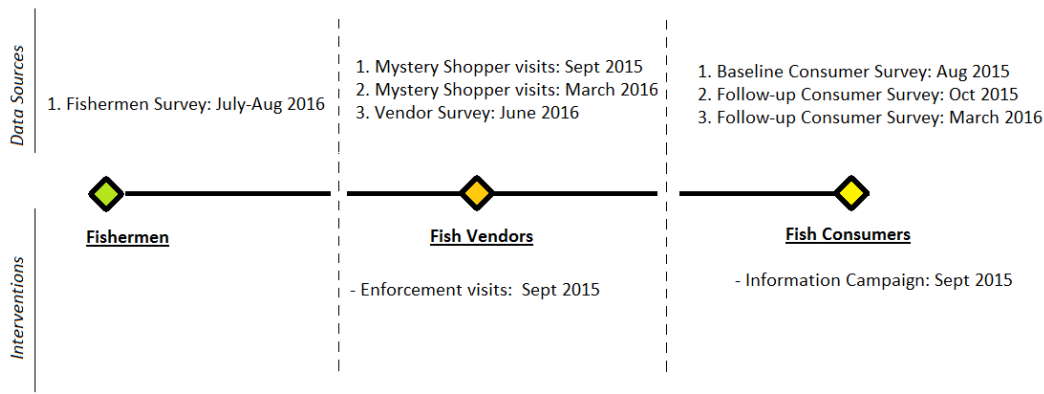

## 5 Results

We start by reporting treatment effects on availability of illegal hake in the market. We then analyze how vendors react in their attempt to circumvent the treatments, and whether our sub-treatment variations limit their ability to do so. We then track market-level spillovers using information from the fishermen and vendor surveys, and re-analyze the main treatment effects accounting for spillovers. Finally, we show effects on prices and quantities of substitutes.

## 5.1 Balance

We did not conduct a full baseline survey, but had access to municipality administrative data and weather data with which we could check balance across treatment arms. The table C.4 in appendix C describes the main variables in each treatment arm.

Overall, the various treatment arms appear well balanced in terms of important socio-economic and weather characteristics (e.g. poverty rate, rainfall). The joint test F-statistics of all variables are insignificant for different treatment arms. The delinquency rate (i.e., per-capita police cases for major offenses) is lower in municipalities assigned to receive the information campaign relative to the control group. We have verified that our main treatment effects presented below are not sensitive to controls for any of the variables that are individually statistically significant in the balance table.

## 5.2 Empirical Strategy

Each market (feria) consists of multiple stalls run by different vendors. The mystery shoppers we employed visited several stalls in each market multiple times during September 2015 and recorded hake fish availability at these stalls. These visits created a stall-day level panel dataset of 906 visits. The first visit to various markets by Sernapesca enforcement officers occurred between Sept 4 and 10, so our panel data consists of 242 visits during the pre-enforcement period that we define as the first week of September, plus 664 visits during the post-enforcement period (the rest of September). We use the following regression specification to evaluate the interventions, where each observation refers to a mystery shopper visit at fish stall $s$, in feria $f$, from circuit $c$ visited on day $t$:

$$y_{sfct} = \beta_0 Post_t + \beta_1 T_c + \beta_2 T_c \times Post_t + \beta_3 y_{sfc0} + X_{ct}'\beta_4 + \varepsilon_{sfct} \qquad (1)$$

$y_{sfct}$ is the outcome variable, such as an indicator for whether illegal hake fish was available at that stall on that day. The treatment assignment ($T_c$) varies at the circuit level . The variable $Post_t$ indicates the post-intervention period, September 8-30.[10] We control for weather on each day, whether the inspector visited the market that day, a few socioeconomic covariates (e.g. municipality crime rate), randomization strata fixed effects, and the baseline (pre-intervention) value of the dependent variable. The error term, $\varepsilon_{sfct}$, is clustered at the circuit level, which was the unit of randomization. The coefficient of interest for the evaluation is the parameter $\beta_2$, which captures the difference between treatment and control groups during the post-intervention period. In most of our tables, we will only report these coefficients, and suppress all others.

### 5.3 Hake Sales Observed by Secret Shoppers

We use the mystery shopper data, where the vendors did not know that their activities were being monitored, to create our main outcome of interest, which is an indicator for whether the vendor was selling hake during the ban period. Column 1 of Table 3 shows the effect of the interventions on whether fresh, visible hake was available for sale in that stall. Column 2 shows effects on whether hake in any form (fresh and visible, hidden in the back, or "frozen" hake that is kept on ice) was available for sale. Each dependent variable is binary, and we report marginal effects from a Probit regression. The three coefficients of interest are the variables with a $\times Post_t$ interaction: which track the effects of the demand-side information campaign, the supply-side Enforcement treatment and the interaction between the two (markets where both supply and demand interventions were simultaneously administered) during the post-intervention period. These three coefficients summarize treatment effects relative to the control group, which is our omitted category.[11]

---

[10]This corresponds to the rollout of the enforcement intervention, and the information campaign letters arrived at households even after this date. There are other reasonable ways to define the post-intervention period, and we make a conservative choice. We have verified that the exact definition of the post intervention period does not affect our main results.

[11]The "Information Campaign" group is a marker for circuits located in municipalities assigned to receive the High-Saturation Information Campaign, where the majority of neighborhoods were treated with the campaign. Appendix Table C.10 explains why we made this modeling choice. Our consumer survey data indicates that the majority (69%) of shoppers we found shopping at *ferias* located in "control" neighborhoods in high-saturation treatment municipalities resided in neighborhoods that were treated. It therefore makes more sense to code such *ferias* as 'treated' with the information campaign. Appendix C.6 shows the results of re-estimating the results in Tables 3, but reverting to coding ferias in control neighborhoods as not treated with information. The results are qualitatively similar.

Table 3: Treatment Effects on Hake Sales

| VARIABLES | (1) Fresh, Visible Hake | (2) Any Hake Available (Hidden, Frozen, Visible) |
|---|---|---|
| Information Campaign Only | 0.080 | 0.029 |
|  | (0.056) | (0.058) |
| Enforcement Only | 0.114 | 0.092 |
|  | (0.070) | (0.060) |
| Information Campaign and Enforcement | 0.078 | 0.100 |
|  | (0.070) | (0.065) |
| Information Campaign Only × Post | -0.133** | -0.131* |
|  | (0.066) | (0.074) |
| Enforcement Only × Post | -0.178** | -0.130 |
|  | (0.082) | (0.089) |
| Info Campaign and Enforcement × Post | -0.179** | -0.139 |
|  | (0.074) | (0.094) |
| Change in Dep. Var. in Control Group During Intervention Period | -0.21 | -0.36 |
| N | 901 | 901 |

This table reports the effect of each treatment arm on the availability of illegal hake fish. The variable Fresh Hake indicates when the hake was available fresh. Hake available indicates when was possible to buy fish in any form. The table reports marginal effects from a Probit regression. Other controls are included: municipality characteristics, strata fixed effects and the average level of the outcome variable in pre-intervention period. Robust standard errors clustered by circuit (the unit of randomization) in parentheses. *** $p<0.01$, ** $p<0.05$, * $p<0.1$

The first three rows indicate that there were no statistically significant differences between treatment and control groups during the pre-intervention period. As expected, significant differences between markets appear after the interventions are launched (after the first week of September). In column 1, vendors in markets exposed to the information campaign are 13.3 percentage points less likely to be selling fresh, visible hake relative to control group vendors. This is quite a large effect, considering that about 43% of vendors in control markets were selling hake during the first week of September, before the interventions were launched. Vendors operating in markets where Sernapesca monitors visit to levy penalties become 17.8 percentage points less likely to sell fresh visible hake. The combination of the two treatments also produces a 17.9 percentage point decrease in hake sales, so there is no evidence that the information campaign complements the enforcement strategy to make it more effective.

When we add "hidden" and "frozen" hake to fresh/visible hake sales, to create a broader dependent variable that captures any type of hake sales in column 2, the treatment effects become smaller and lose statistical significance. Taken together, the two columns suggest that while the interventions reduced vendors' propensity to engage in illegal activity visibly and openly, some of those vendors substituted away to hiding the fish, or putting it on ice, but still continuing to sell hake. The magnitude of the *reduction* in the treatment effect from column 1 to 2 is larger for the enforcement treatment arms. In summary, while our treatments were successful in reducing the type of illegal activity that could be easily monitored by regulators (visible sales in column 1), it is not so clear whether it actually reduced the underlying environmental harm that we care about (column 2). The difference (possibly) stems from the defensive strategies that vendors adopt in response to the regulators' monitoring attempts. We will explore those defensive strategies in greater detail in section 5.5.

## 5.4 Consumer Behavior

We consider the mystery shopper data to provide the most reliable measure of the illegal behavior we track. Nevertheless, we also directly surveyed consumers at markets about their purchase behavior. This allows us to report results on the (self-declared) consumption of fish using consumer surveys conducted before and after the ban. The first column of Table 4 shows treatment effects on the number of times that consumers interviewed at the market report buying hake fish during the previous month. We see significant decreases in (self-reported) hake purchase across all treatment arms, and so results are generally consistent with the mystery shopper survey. However, in these consumer reports, the treatment effects appear larger in information campaign areas (relative to enforcement areas). This may be because the consumers received direct communication in the information areas, which may create some self-reporting bias.

Consumer behavior was also indirectly influenced by the enforcement activity. Not only did self-reported hake purchases decrease there relative to control markets, the third column also shows that consumers were about twice as likely (or 8-11 percentage points more likely) to mention to our enumerators, totally unprompted, that they did not buy hake fish because there was a September ban in place. Our enumerators did not specifically ask consumers any questions that mentioned the ban, but were instructed to note down whenever a consumer spontaneously mentioned the ban. Consumers treated with the information campaign were 15 percentage points more likely to mention the September ban unprompted, so evidently the treatments were at least successful in spreading more information and awareness relative to control areas.

Table 4: Treatment Effects on Fish Consumption

| VARIABLES | (1) Num. Times Hake Purchased | (2) Mention Ban (unprompted) |
|---|---|---|
| Information Campaign Only | -0.586*** | 0.147*** |
|  | (0.181) | (0.046) |
| Enforcement Only | -0.238** | 0.083* |
|  | (0.096) | (0.047) |
| Info Campaign and Enforcement | -0.208** | 0.108** |
|  | (0.093) | (0.052) |
|  |  |  |
| Mean Dep Var Control Group | 0.37 | 0.10 |
| N | 3218 | 3319 |

This table presents the effect of different treatments on the reported consumption of hake fish during September 2015. The column 1 shows the marginal effects from a Poisson regression because the dependent variable is count data, the column 2 shows marginal effects from a Probit regression. Consumers were not asked about the ban, but surveyors registered if the ban was mentioned spontaneously. These regressions include socioeconomic characteristics and strata fixed effects. The Standard errors are clustered based on the circuit where the survey was collected. Robust standard errors clustered by circuit in parentheses.*** p<0.01, ** p<0.05, * p<0.1

### 5.5 Vendor Reactions

We quickly learnt from our mystery shoppers that vendors undertake two major defensive strategies in an effort to continue selling illegal hake in September: they hide the hake, or they put the fish over ice and claim that it was caught legally in August, and frozen since then.

*Hiding:* We devised the mystery shopper data collection strategy for this project because they are in principle able to observe a lot more than government inspectors. For example, even if hake was not visibly on sale in the stall (which our mystery shoppers later noted down after completion of the visit), they were able to ask vendors whether there was any hake available for sale. The mystery shoppers noted down each occurrence of "hidden hake" for us, but we never shared the specific vendor or feria identity with our government partners, to protect vendor privacy and abide by our research ethics (IRB) protocol. These data were very useful for the evaluation, but were never used to target enforcement.

Our mystery shoppers observed the practice of hiding in 3-4% of cases in treatment areas. The "hidden" hake fish was often stored in a cooler behind the board that displayed the stall's fish prices. This was clearly used as a strategy to circumvent the September ban: We conducted another mystery shopper survey six months after the ban, and there we did not observe even a single stall selling fish that's not publicly-visible. Hiding fish is costly for vendors, because displaying the fish available for sale on any given day and shouting out to potential customers are the main marketing tools at the vendors' disposal. Many of our mystery shopper noted down in the qualitative comments sections of the survey instrument that they observed regular consumers asking vendors for hake when it was not visible. The hiding strategy evidently works because many consumers are willing to partake.

*Freezing:* On paper, vendors are not allowed to sell hake fish in any form in September. In practice, Sernapesca inspectors were more lenient with vendors who were detected selling "frozen" hake. This is the practice of putting the fish on ice and claiming that it was harvested in August, before the ban. Unlike "hiding", we had not anticipated this reaction, but a couple of our mystery shoppers noted the practice for us early enough, such that we were able to collect systematic data on it. Comparing our mystery shopper reports with administrative data from Sernapesca on fines levied based on the registry of inspector visits suggests that inspectors were much less likely to levy penalties when the vendor was claiming to sell "frozen" hake.

Selling frozen hake is quite costly for vendors because consumers prefer the taste of fresh fish, and because freezing requires freezers and access to electricity. Using our other rounds of data, we see that freezing is virtually non-existent during the rest of the year. So this does appear to be a strategy that vendors use to circumvent the September ban.

There are several pieces of circumstantial evidence in our data that this is all pretense; that fishermen and vendors are not actually protecting the environment by catching fish in August and freezing it until September. First, we document more freezing in the second half of September 2015 than the first half (especially in markets inspectors visited on a predictable schedule), after vendors have had a chance to learn about the enhanced regulatory activities that year. Real freezing would have been much less costly to engage in during the first half of the month. Second, we collected data on stall characteristics, and availability of a freezer in a stall is not at all predictive of freezing. If anything, our mystery shoppers find that stalls without freezers are more likely to be selling frozen fish post-intervention. Third, many secret shoppers noted down that in their conversations with vendors, many vendors admitted (and even insisted) that the fish was fresh even though it was labeled as frozen.

Figure 3 shows the prevalence of freezing and hiding across treatment groups. We divide up the control group into markets that have another circuit that is randomly assigned to enforcement within 10 kilometers (to capture spillover effects), and *pure control* markets that are at least 10km away from any treated area. Several notable patterns emerge:

1. We do not observe any hiding or freezing at all in pure control markets. In contrast, 7.2% vendors operating in circuits that received Sernapesca inspector visits sell frozen fish (p-value 0.00), and 3.2% of those vendors engage in hidden hake sales (p-value 0.01).

2. Vendors operating in circuits exposed only to the information campaign did not engage in any hiding or freezing at all. It appears that vendors employ these defensive strategies only against Sernapesca inspectors, not informed consumers.

3. 4% of vendors who operate in control markets - but located close to treated areas - engage in hiding and freezing, in contrast to 0% in *pure control* markets (p-value 0.02). There appear to be some spatial spillovers in information about Sernapesca visits, and in vendor behavior. We will explore the nature of these spillovers at greater depth in Section 6.

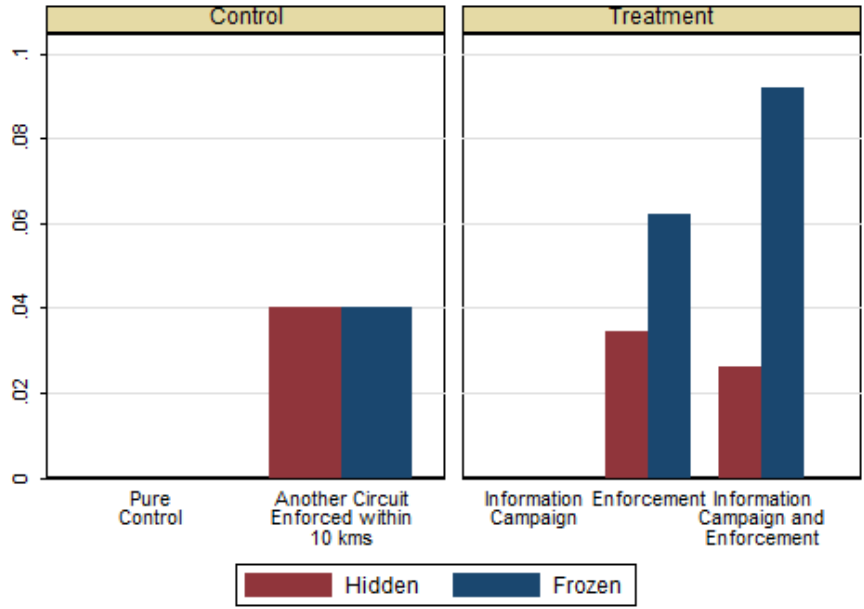

Figure 3: Hidden and Frozen Hake Fish

This figure shows the unconditional mean of hidden hake for different treatment status. The level of frozen hake is statistically different from zero for markets assigned to Enforcement and Enforcement and Info Campaign. The level of hidden is statistically different from zero for markets with Enforcement and spill-overs.

The proportion of vendors selling hake that sell the "frozen" variety increases week-to-week all through September. Only 1.5% of sellers sell frozen hake at the beginning of September, 16% during Sept 3-9, 53% during the following week, continually increasing to 71% during the last week of September. Evidently vendors learn how to employ this strategy over the course of the month.

These hiding and freezing reactions suggest that vendors learn about Sernapesca inspector visits and penalties, react to them, and adapt their selling strategies to circumvent the enforcement. It is important to take such reactions into account when we are interested in comprehensively evaluating the longer-term consequences of policy. In our specific case, the enforcement strategy produced larger decreases in fresh, visible hake sales compared to the information campaign (column 1 of table 3), but once these vendor adaptations are taken into account, the relative policy ranking reverses (column 2 of table 3). We next explore whether the specific design of the enforcement strategy can be altered to make enforcement more effective, accounting for the (anticipated and unanticipated) defensive strategies that vendors may employ. The data analysis will also allow us to explore how and what agents learn about the regulator's activities.

## 5.6 Variations in the Design of the Enforcement Strategy

Varying the enforcement strategy involves experimentally manipulating the schedule of visits in two dimensions: Predictability and Frequency. Table 5 uses the mystery shopper data, and repeats the regression setup of Table 3, except that the enforcement treatment is now sub-divided into areas where the monitoring schedule was either predictable or unpredictable (panel A), or sub-divided into areas where monitoring was conducted at high versus low frequency (panel B).

Table 5: Treatment Effect on Hake Sales by Enforcement Strategy

| VARIABLES | (1) Fresh, Visible Hake | (2) Any Hake Available (Fresh-Visible, Hidden or Frozen) |
|---|---|---|
| Panel A: Variation in Predictability of Enforcement Schedule | | |
| Information Campaign only | -0.122* | -0.134* |
|  | (0.068) | (0.073) |
| Enforcement on Predictable Schedule | -0.092 | -0.060 |
|  | (0.069) | (0.083) |
| Enforcement on Unpredictable Schedule | -0.243*** | -0.192** |
|  | (0.086) | (0.094) |
| Panel B: Variation in Frequency of Enforcement Visits | | |
| Information Campaign only | -0.123* | -0.135* |
|  | (0.067) | (0.072) |
| High Intensity Enforcement | -0.063 | -0.070 |
|  | (0.086) | (0.095) |
| Low Intensity Enforcement | -0.198** | -0.162* |
|  | (0.083) | (0.090) |
| | | |
| Change in Dep Var in Control | | |
| During Intervention | -0.21 | -0.36 |
| N | 901 | 901 |

This table presents the coefficient corresponding to the interaction term $T_c \times Post_t$ for each treatment. To retain statistical power, the cells "Enforcement only" and "Enforcement + Info Campaign" from Table 3 are combined under "Enforcement" and then sub-divided by schedule predictability (panel A), or intensity (panel B). So these coefficients should be interpreted as the average effects of enforcement when half the sample is also exposed to the information campaign. Panel A includes a dummy for the intensity sub-treatment, and Panel B includes a dummy for the predictability sub-treatment, but those coefficients are not shown. Each regression controls for the dependent variable in pre-intervention period, strata fixed effects and municipality characteristics. Probit regression marginal effects are reported. Robust standard errors clustered by circuit in parentheses.*** p<0.01, ** p<0.05, * p<0.1

Panel A shows that the enforcement strategy was generally more effective when the schedule was made unpredictable. When enforcement follows a predictable schedule (e.g. every Tuesday at 10am), its effect is not statistically different than zero. However, when we make the monitoring visits difficult for vendors to predict, we see two important changes: (a) There is a much larger and statistically significant decrease of 24% in fresh, visible hake sales, and (b) There remains a statistically significant decrease of 19% in the availability of any type of hake even after we account for vendor defensive reactions like hiding and freezing the hake. In other words, the lack of predictability also apparently makes it difficult for vendors to effectively engage in strategies like hiding and freezing. We will examine this directly below using data on the prevalence of hiding and freezing across these sub-treatment arms.

Panel B shows results separately for the subgroup of vendors who received monitoring visits once a week (low intensity), and other vendors who were visited twice a week, which means that monitors followed a circuit around in the different market locations where those vendors set up stalls on different days of the week (high intensity). The high intensity visits limit opportunities for spatial and temporal displacement of illegal hake sales, so we had expected it to work better.

The results in Panel B show that the strategy backfired, and that our expectations were incorrect. Enforcement is more effective at reducing hake sales in markets that were visited less frequently. One possible explanation is that more frequent interactions with the inspectors allow vendors to learn about the loopholes of the enforcement strategy more quickly, and how to circumvent it. We explore this in figure 4 using data on the prevalence of "hidden" hake sales. Hiding is about 3.5 times as common under the high-frequency/predictable enforcement regime than in every other cell (p-value 0.01). Making many monitoring visits on a predictable schedule allows vendors to learn most about the monitoring schedule, and evidently makes it much easier for them to engage in adaptive strategies that help them circumvent the enforcement efforts.

Figure 4: Hidden Hake by Enforcement Schedule

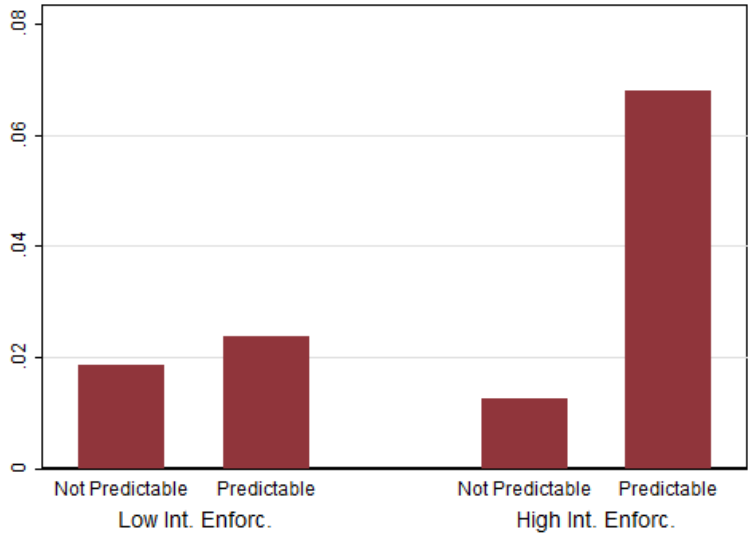

This figure shows the unconditional mean of hidden hake for different sub-groups of enforcement for post-intervention period. The difference between the fourth column and the rest is statistically significant at 1%.

Both hiding *and* freezing are more common when we follow a predictable monitoring schedule. To investigate whether vendors "learn" to cheat after they experience monitoring visits, Figure 5 plots the incidence of freezing separately for the first half of September and the second half. In markets with no enforcement activity, the incidence of frozen fish drops from the first to the second half of the month. This would be consistent with the biology and economics of actual freezing - it's easier and cheaper to freeze August-caught fish for 7 days rather than 20 days. However, in markets inspectors monitored on a predictable schedule, not only is the incidence of freezing higher than under no-monitoring or under unpredictable monitoring, freezing becomes more common in the second half of the month (10%) than in the first half (7.2%).

Figure 5: Frozen Hake by Schedule of Enforcement

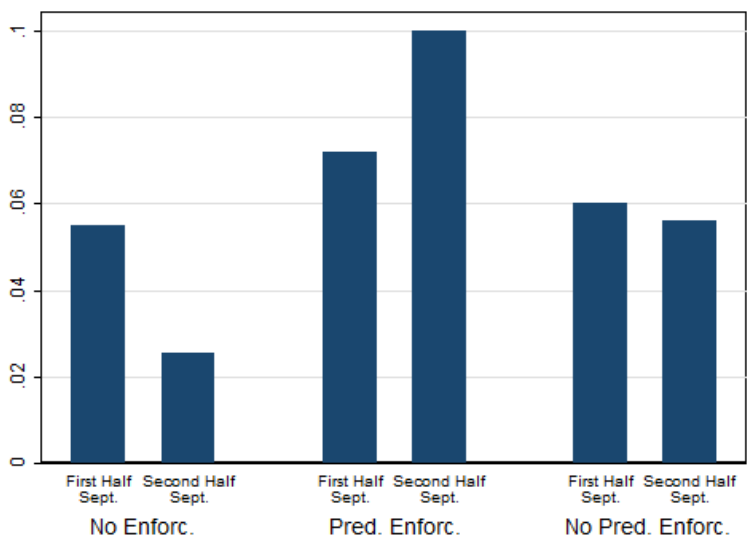

This figure shows the unconditional mean of frozen hake for different sub-groups of enforcement for the first and the second half of the month. Markets with predictable enforcement increased the proportion of frozen hake over time.

In summary, we see evidence that it is easier for vendors to cheat the inspectors under predictable monitoring, and there is some weak evidence that learning how to cheat over time is easier when the monitoring schedule is predictable. All of this results in enforcement being much more effective at reducing illegal hake sales - even after accounting for the cheating - when the inspectors visit to levy fines using a random, unpredictable schedule.

## 5.7 Change in Number of Stalls Selling Fish

Even though most fish vendors sell multiple types of fish, one possible effect of our intervention is that some fish vendors exit the market altogether during September, when they find it difficult to sell hake. This would lead to selection in whom we observe in our follow-up data. Figure 6 shows that the average number of fish stalls does decrease in the markets randomly assigned to the enforcement treatment, especially during the second half of September. This itself is an important effect of the treatment, but it also changes the interpretation of the treatment effect on the propensity to sell hake reported elsewhere in the paper. This finding indicates that some vendors facing enforcement drop to zero sales, but this is not captured in Table 3. That makes the coefficients reported in that table possibly under-estimates of the true effects.

How large an under-estimate it is depends on how likely it is that those "missing" stalls would otherwise be selling hake. If those stalls exit from markets where hake is not usually sold anyway, then correcting our estimates for these "missing" stalls would not change our results substantially. Table C.7 in Appendix C.3 describes how we correct our estimates for stalls exiting. The correction makes the effect of Enforcement larger than that reported in Table 3, but it does not affect the coefficients for other treatments very much.

Figure 6: Number of stalls in Feria by Treatment Assignment

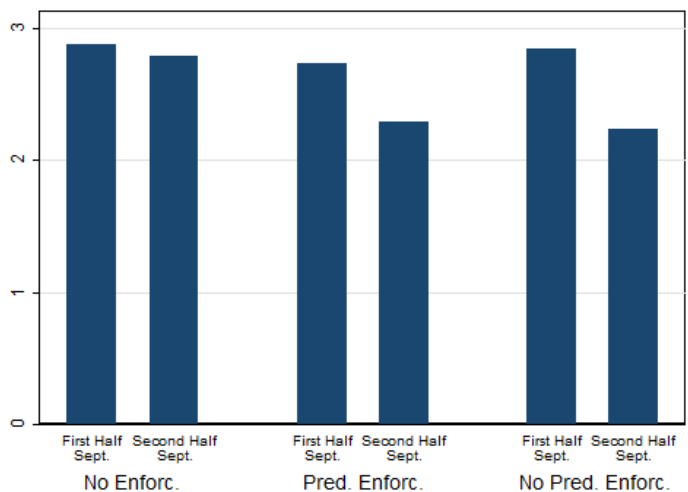

This figure shows the average number of stalls of ferias visited by mystery shoppers in the first and the second half of the month for different treatment arms. Markets assigned to receive enforcement showed a decrease in the number of stalls between the first and the second half of the month.

# 6 Spillovers and Market Level Effects

While our experiment was targeted to reduce hake sales in treated *ferias*, it may have had spillover effects on control markets through information transmission, or by changing equilibrium prices (Blattman et al., 2017). It may also have affected the behaviors of other market actors, such as the fishermen who supply to vendors. It could have also changed the prices and quantities of other fish that can act as substitutes for hake. We collected additional data to study these spillovers and equilibrium effects, including a survey of fishermen, a survey of vendors to understand their social and supply-chain connections to vendors operating in other markets, GIS data on the location of all markets, and data on the prices of hake and other substitute fish. The vendor and fishermen surveys allow us to map the supply chain for all ferias. The spatial dispersion of *ferias* where vendors sell and *caletas* where the fishermen bring in their catch along the long Chilean coast, creates large variation in geographic and social connections that are useful for detecting spillovers.

## 6.1 Spillovers on Control Markets

We identified three primary channels through which treatment may affect behavior of control markets, and collected data on each channel:

1. *Spatial spillover:* Control markets located geographically close to a treated market may feel the effects of treatment because they share consumers with the treated area.

2. *Social spillover:* If control market vendors are socially connected to vendors operating in treatment areas, they may be more likely to learn about *Sernapesca*'s enforcement activities.

3. *Supply chain spillover:* Treatment and control vendors may source from the same fishermen. If a supplier changes fishing behavior due to treatment, that could indirectly affect fish sales in control markets.

Of these different channels, econometrically, an increase in fish sales in control markets due to changes in either equilibrium price or behavior by any actor in the supply chain is of greatest concern. Examples include an increase in equilibrium hake price if the supply curve for hake shifts left due to enforcement, or fishermen dumping all excess hake in control markets when vendors in treated markets are unwilling to buy hake. This is because under those circumstances, a treatment-control difference will appear to show that the treatment was effective, when in fact hake sales were simply spatially displaced towards the control group. Our regressions would over-estimate the effects of treatment in that scenario. This is why it's important for us to re-investigate these effects controlling for these sources of spillovers.

We follow a procedure similar to Miguel and Kremer (2004) in estimating treatment effects in the presence of spillovers. We divide the control markets into subgroups; (a) Control areas that are more likely to have been affected by treatment due to geographic or social or supply chain connections, which we call "Spillover Group", and (b) Control areas un-connected to treatment markets, which we call "Pure Control".

In Table 6, we re-estimate our main results on the effects of predictable and un-predictable enforcement originally reported in Table 5, but now controlling for potential channels of spillover effects.[12] The first column presents results without spillover variables, and shows that relative to all other markets, unpredictable enforcement reduces hake sales by 15.7 percentage points. The second column controls for spatial spillovers, with the indicator "within 10km of Treated Market" turning on for untreated markets that have at least one treated *feria* within a distance of 10 kilometers. The coefficient of this variable is negative but small, suggesting very limited spillovers based on shared consumers due to geographic proximity. The third column includes an indicator for control markets where at least one vendor reported that they knew a vendor in a different market that was assigned to the enforcement treatment. The coefficient on this variable suggests that there was a 7 percentage reduction in hake sales due to this "social spillover" (similar in magnitude to the effect of predictable enforcement), but this effect cannot be statistically distinguished from a zero effect with much confidence. Finally, column 4 includes an indicator for control markets who source from fishermen operating out of *caletas* that primarily supply to other markets that were assigned to the enforcement treatment. We again see a 7.7 percentage point reduction in hake sales in control markets that are connected to treated markets through shared suppliers, but the effect is not statistically precise.[13]

Importantly, accounting for these spillover effects make the main treatment effects of unpredictable enforcement on enforced areas a little larger and more statistically precise. This is because controlling for spillovers allow us to compare treated areas to the subset of "pure" control areas unaffected by the treatment. Relative to such pure control markets, the unpredictable enforcement reduces hake sales by 18-20 percentage points (as opposed to 16 pp estimated in col. 1).

---

[12] Note that sub-dividing the control group into "pure control" and "spillover markets" reduces the number of markets allocated to the omitted category. To retain sufficient statistical power, we therefore focus on re-estimating the effects of enforcement treatment variations only, because spillovers cause the greatest econometric concern (of over-estimating effects) for this particular result. In contrast, spillovers from the information treatment should enhance the effects of treatment. In this setup, some of the markets in the omitted category received the information treatment, so the regression coefficients will look a little smaller in this table compared to Table 5.

[13] Vendors connected to a larger number of other circuits are more prone to being exposed to the treatment, and that variation is not random. To control for this, we include a full set of dummy variables for the number of other circuits that each reference circuit is connected to, separately for spatial, social and supply-chain connections. Thus, the variation of exposure to spillovers stems only from the treatment status of other markets, which is exogenous because it was randomly assigned. (Miguel and Kremer, 2004).

Table 6: Treatment Effects on Hake Sales Controlling for Spillovers to Control Markets

| VARIABLES | (1) Any Hake (Fresh/Hidden/Frozen) | (2) Any Hake (Fresh/Hidden/Frozen) | (3) Any Hake (Fresh/Hidden/Frozen) | (4) Any Hake (Fresh/Hidden/Frozen) |
|---|---|---|---|---|
| Enforcement on Predictable Schedule | -0.023 | -0.030 | -0.076 | -0.058 |
|  | (0.083) | (0.069) | (0.080) | (0.060) |
| Enforcement on Unpredictable Schedule | -0.157* | -0.167* | -0.199** | -0.177** |
|  | (0.091) | (0.075) | (0.084) | (0.084) |
| Spatial Spillover |  | -0.017 |  |  |
| (within 10 km of Treated market) |  | (0.082) |  |  |
| Social Connection Spill-over |  |  | -0.071 |  |
| (Vendor knows a Treated Vendor) |  |  | (0.076) |  |
| Supply-Chain Spill-over |  |  |  | -0.077 |
| (Sources from same *Caleta* as Treated Vendor) |  |  |  | (0.081) |
| Change in Dep Var in Control During Intervention | -0.36 | -0.36 | -0.36 | -0.36 |
| N | 901 | 901 | 901 | 901 |

This table re-estimates treatment effects controlling for possible spillover effects from treatment to control markets. We only present the coefficient corresponding to the interaction term $T_c \times Post_t$ for each treatment. Controls for $T_c$, $Post_t$, covariates, and baseline value of the dependent variable are included, but those coefficients are not shown. The table reports marginal effects from a Probit regression. The dependent variable is an indicator for any type of hake (fresh-visible, hidden or frozen) for sale in the stall. Robust standard errors are clustered by circuit, which was the unit of randomization. *** p<0.01, ** p<0.05, * p<0.1

## 6.2 Treatment Effect Transmission along the Supply Chain

For the supply chain spillover channel to be relevant, the fishermen supplying hake to these vendors must have altered their behavior in some way. To understand those changes, we directly survey fishermen operating out of every *caleta* (fishing village) that serves the markets in our sample.[14] The reactions of fishermen are particularly important to track because our interventions conducted at the final point-of-sale has to somehow get transmitted up the supply chain to fishermen, for these interventions to ultimately protect the hake population. Only if fishermen start perceiving the effects of these interventions on demand conditions will they change fishing behavior in ways that improve environmental outcomes that the policymaker cares most about.

Since we did not have baseline data from fishermen for years preceding the September 2015 ban, we ask them retrospective questions in 2016, in which the fishermen are asked to compare demand and profits during September 2015 (when our interventions were launched) relative to September 2014. To minimize possible response bias given the government fishing ban, we were careful to phrase our questions generically, to cover revenues earned from all types of fish, and not just hake specifically. To report treatment effects on fishermen, we have to connect each *caleta* to treatment and control markets. We use the vendor survey on the structure of the supply chain -i.e. which caletas each vendor buys from - to link fishermen to the randomized treatments.

---

[14] A few caletas in the regions covered by our sampling frame are only used by divers who harvest seafood, not fish -and we therefore exclude those *caletas*.

Table 7 reports results. Column 1 shows that fishermen operating out of *caletas* that sell to at least one circuit which had been randomly assigned to enforcement, are 32 percentage points more likely to report that they earned less in September 2015 compared to September 2014, relative to fishermen in caletas that supply to control group *ferias*. Fishermen operating out of caletas that supply to both enforced markets and to markets that experienced the information campaign were 45 percentage points more likely to report lower revenues during the month of the interventions, compared to the same month in the previous year. In column 2, these fishermen are also significantly more likely to report that vendors were less willing to buy hake in September 2015 compared to the previous year. So our interventions not only affected the behavior of vendors, as observed by our mystery shoppers, those changes in behavior were also perceived by fishermen further upstream in the fish supply chain. Column 3 shows that fishermen linked to the information campaign areas are more likely to report that final consumers are aware of the hake ban. Information clearly traveled up and down the supply chain.

Table 7: Treatment Effect Transmission to Fishermen in Caletas

| VARIABLES | (1) Earned Less in Sept 15 than Sept 14 | (2) Feria Vendors buy less Hake in Sep15 compared to Sept 14 | (3) Consumers are informed of Hake Ban |
|---|---|---|---|
| At least one circuit Enforced | 0.323*** | 0.195 | 0.009 |
|  | (0.105) | (0.303) | (0.161) |
| Info Campaign | 0.120 | -0.049 | 0.364* |
|  | (0.158) | (0.334) | (0.192) |
| At least one circuit Enforced and Info Campaign | 0.451*** | 0.626* | 0.217 |
|  | (0.131) | (0.329) | (0.201) |
|  |  |  |  |
| Mean Dep Var Control Group | 0.31 | 0.40 | 0.77 |
| N | 202 | 179 | 217 |

This table reports OLS coefficients based on fishermen responses. The variable Information campaign correspond to caletas located in municipalities assigned to receive any level of information campaign. The variable "At least one circuit enforced" considers all circuits located in the same municipality of the caleta. Socioeconomic variables of the caletas are included as covariates. In average three fishermen were surveyed in each caleta. Robust standard errors clustered at caleta level in parentheses. *** p<0.01, ** p<0.05, * p<0.1

## 6.3 Market Level Effects

In this section we attempt to describe how the market for fish in the aggregate may have been affected by our treatments. We tried to collect data on prices, and the availability of other fish species in the same markets where hake is sold. The September ban is only specific to hake fish, so we might expect consumers to substitute to other fish varieties. This may be because informed consumers choose to avoid hake fish during the ban, or because the enforcement treatment reduces hake availability or increases its price.

The universe of data from all markets suggests that there are seven possible fish substitutes for hake,[15] but a typical stall only offers two or three varieties of fish. Table C.2 in the Appendix describes the availability and price of different fish species observed by mystery shoppers in ferias during September 2015. The most common fish substitute is pomfret, which can be found in two-thirds of all markets. Pomfret is larger and (arguably) more tasty than hake fish. In Table 8, we study the availability of pompfret (column 1), or any other non-hake fish including pomfret (column 2), as a function of the treatment status of the market where the fish stall is located.

Table 8: Do Vendors Substitute to Selling Other Fish in Response to Treatment?

|  | (1) | (2) |
| --- | --- | --- |
| VARIABLES | Pomfret Available | Any Other Fish Available |
| Information Campaign Only | 0.146 | 0.004 |
|  | (0.098) | (0.035) |
| Enforcement on Predictable Schedule | 0.133* | 0.027 |
|  | (0.079) | (0.031) |
| Enforcement on Unpredictable Schedule | 0.115 | 0.065* |
|  | (0.078) | (0.033) |
|  |  |  |
| Change in Dep Var in Control Markets |  |  |
| During Intervention | 0.29 | 0.09 |
| N | 901 | 6328 |

The table reports marginal effects from a Probit regression. The unit of observation in the first column is stall x secret shopper visit, and in the second column is stall × secret shopper visit × possible substitute fish variety. We only present the coefficient corresponding to the interaction term $T_c \times Post_t$ for each treatment. Controls for $T_c, Post_t$, covariates, and baseline value of the dependent variable are included, but those coefficients are not shown. Robust standard errors are clustered by circuit in parentheses. *** p<0.01, ** p<0.05, * p<0.1

---

[15]The variety of substitutes depends on the latitude of the market. The substitutes most commonly found by mystery shoppers in ferias are: pomfret, mackerel, silverside, salmon, sawfish, albacore and southern hake.

The penultimate row of the table indicates that stalls in control markets are 29 percentage points more likely to start selling pomfret during the September hake ban, so it appears that vendors in general move towards substitutes during the ban. The increase in pomfret sales during September is larger in treated areas (by a further 12-15 percentage points, which results in a 41-44 percentage point increase during the hake ban), but the treatment-control differences are barely statistically significant. The p-value for only one of the three coefficients (associated with Predictable Enforcement) is below 0.10. Column 2 investigates treatment effects on the vendor's decision to offer each of seven different fish substitutes for hake. The sample size is larger in this regression because selling each fish variety is treated as a separate decision, but our standard errors are still clustered by the unit of randomization of the treatment (the circuit). The coefficients indicate that vendors who faced unpredictable enforcement become 15.5 percentage more likely to switch to selling other fish during the hake ban, compared to the 9 percentage point increase in control markets. This 6.5 percentage point treatment-control difference is statistically significant (p=0.051).

## 6.4 Effects on Prices

We collected data on fish prices during all our mystery shopper visits. However, prices are observed only when the fish is available for sale. Indeed, during September, hake is only available in 26% of markets, which implies that the price of fish can be analyzed using a relatively small sample of observations. Further, our earlier analysis indicates that treatment changes the propensity to sell illegal hake fish. In other words, treatment affects the selection of which prices are observed. There are therefore large sample-selection issues that complicates any analysis of treatment effects on prices, and we refrain from running regressions on the price of hake. The most consumed fish during September (and second most consumed fish during the rest of the year) is Pomfret, which is available in 68% of the stalls (see Appendix Table C.2). Since pomfret is more often available (and not banned), we instead run regressions to study treatment effects on the price of pomfret.

As a descriptive exercise, Figure 7 shows that the price of hake increased week-to-week in September, over the course of the ban period. Pomfret prices fell by 10% in the second week and that lower price remained stable thereafter. This time-series pattern in prices is consistent with fishermen upstream in the supply chain shifting away from hake and towards catching pomfret during our interventions in September 2015. Through conversations with fishermen during our survey, we learnt that they are able to adjust their fishing strategy to target different species if there are market signals that hake demand is low. To do so, they change the location and depth at which their nets are dropped.

Figure 7: Log Prices of Fish During the Ban

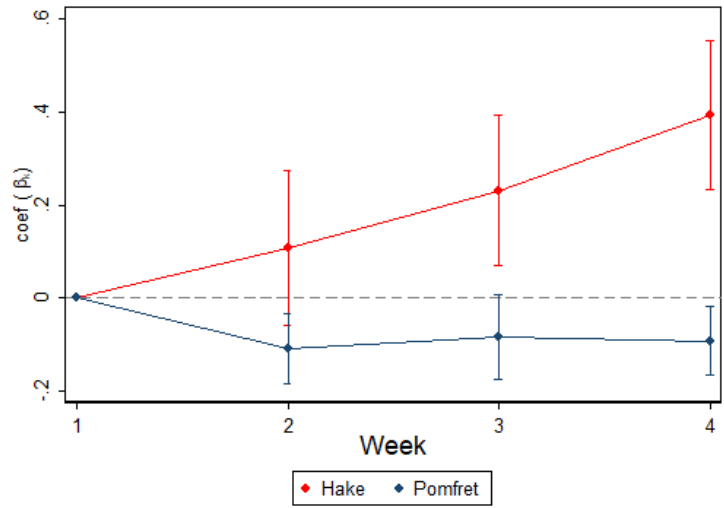

This figure shows the evolution of log prices for hake and pomfret, using the first week as a reference. The price of hake continuously increased over the course of September 2015. Hake was 40% more expensive by the fourth week relative to the first week. The price of pomfret decreased around 10% after the first week.

Table 9 shows pomfret prices observed in a stall as a function of our randomized treatments assigned to the market where that stall is (column 1), as well as the price of any other substitute fish including pomfret (column 2). We find that the price of substitutes increase as a result of the information campaign (that discouraged hake consumption), suggesting that part of the demand for hake shifted towards substitutes. Relative to the control group, markets that received enforcement show small and insignificant price decrease. The fact that we observe such price effects suggests that fish markets are at least somewhat segmented.

Table 9: Treatment Effect on Fish Prices

|  | (1) | (2) |
|---|---|---|
| VARIABLES | Log Price Pomfret | Log Price Substitute |
| Information Campaign Only | 0.210* | 0.140 |
|  | (0.109) | (0.096) |
| Enforcement Only | -0.017 | -0.021 |
|  | (0.066) | (0.055) |
| Info Campaign and Enforcement | 0.081 | 0.047 |
|  | (0.065) | (0.059) |
| Change in Dep Var in Control |  |  |
| During Intervention | -0.20 | -0.27 |
| N | 614 | 939 |

The table reports treatment effects on hake substitutes' price from OLS regressions. The outcome variable is the log of price per kilo. The unit of observation in the first column is *stall with pomfret available × secret shopper visit*, and in the second column is *stall with any substitute available × secret shopper visit × substitute available fish variety*. We only present the coefficient corresponding to the interaction term $T_c \times Post_t$ for each treatment. Controls for $T_c$, $Post_t$, covariates, and baseline value of the dependent variable are included, but those coefficients are not shown. Robust standard errors are clustered by circuit in parentheses. *** p<0.01, ** p<0.05, * p<0.1

## 7 Relative Cost-Effectiveness of Enforcement vs. Information

We conducted an information campaign in addition to the enforcement activities because our regulatory agency partner, *Sernapesca*, believed that demand-side strategies, if they work, would be more cost-effective and easier for them to implement in the future. Given the complications associated with enforcing regulations documented in this paper, and the complexity of designing regulations that are robust to unanticipated defensive reactions from enforced agents, it is useful to determine how cost-effective the enforcement strategies were relative to an information campaign. We collected data from *Sernapesca* on the full administrative costs of implementing each treatment, so that we can report on the relative cost-effectiveness of enforcement and information strategies. This allows us to report on the best use of (limited) public resources to protect hake populations.

We define effectiveness of our interventions on the basis of our data on the reduced probability of observing hake sales in treated ferias, accounting for hidden and frozen fish. Since the fish sold in ferias comes directly from fishermen villages and was harvested the same day or the day before, we assume that reduced hake sales directly correlates with the decrease in hake fishing in caletas. This assumption is supported by the fact that our interventions were conducted at scale covering all major markets where hake is sold in the sampling regions, This means that our data are net of "leakages" of hake from our sampling areas. The fishermen survey results we report in Section 6.2 also suggests that fishermen did feel the effects of the interventions.

Table 10: Cost-Effectiveness Analysis

|  | (1) Reduction of Hake Sale | (2) Units of Hake Saved | (3) Implementation Costs (USD) | (4) Cost of Saving One Hake (USD) |
|---|---|---|---|---|
| Enforcement |  |  |  |  |
| *Overall* | 0.13 | 10,399 | $ 62,900.25 | $ 6.05 |
| *Unpredictable* | 0.192 | 15,358 | $ 69,190.27 | $ 4.51 |
| *Low Intensity* | 0.162 | 12,959 | $ 53,475.84 | $ 4.13 |
| Info Campaign | 0.13 | 3,257 | $ 16,213.53 | $ 4.98 |

This table shows the benefits and costs of implementing each intervention. Column (1) reports the estimated effects (in percentage points) of treatments in the sale of any type of hake. Column (2) is computed based on the numbers of stall per feria, number of days a week the feria operate and number of fish available in a normal stall. Column (3) is reported by Sernapesca and represents a combination of fixed and variable costs. Finally, column (4) correspond to the ration of (3) over (2).

In Table 10, we conduct the relative cost-effectiveness analysis by taking our best estimates of the effects of treatments on reduction in hake sales and combining it with an estimate of the number of fish available in the market that we compute using the data we collected from vendors. This allows us to create an estimate of the extra hake fish that are "saved" due to these treatments.

We compare this number with the cost of implementing each intervention to compute how much it cost to save each fish, under the different assignments. Sernapesca's prior was correct that overall, the information campaign appears more cost effective than the enforcement strategy. This is partly because enforcement becomes less effective as vendors learn to hide and freeze fish and circumvent regulation. Enforcement costs US$6.05 per saved fish, compared to $4.98 under the information campaign.[16]

---

[16] More details on the inputs used in these calculations are provided in section D in the Appendix.

However, once we examine specific versions of the enforcement strategy that were more successful at curbing hake sales, we see that sending monitors on an unpredictable schedule is an even more cost effective way to protect hake, even after accounting for the fact that unpredictable monitoring schedules were much more costly for *Sernapesca* to maintain because it required slack personnel capacity. The cost of "saving" a hake via unpredictable enforcement drops to $4.51. Not surprisingly, low-intensity enforcement (i.e. a less frequent monitoring schedule) is most cost-effective (only $4.13 per saved hake) because it was both more effective at reducing hake sale than high-intensity enforcement, and it was obviously also cheaper to implement.

These calculations are useful to gauge the *relative* cost-effectiveness of alternative strategies to protect hake, but it does not tell us whether any of these strategies would pass a cost-benefit test. Sophisticated benefit calculation would require us to take a stance on the biology of hake fish (how saving a hake in September 2015 translates into a dynamic effect on the hake population via reproduction), and the ecological value of protecting hake. These considerations are outside the scope of our analysis, but our results can be easily combined with benefit numbers from ecology studies. The analysis in this paper takes the government's regulatory goal ("Protect hake fish") as given, and studies the consequences of enforcing that regulation, and analyzes the best ways to achieve that goal.

## 8 Conclusion

Research in many fields of applied microeconomics evaluate the effects of new regulations, such as anti-corruption campaigns, fines levied on agents who do not conform to health or hygiene or environmental standards, or penalties for tax evaders. The effectiveness of such policies depend on the (sometimes unanticipated) reactions of the regulated agents to the new enforcement regime, which is in essence a micro version of the "Lucas critique" (Lucas, 1976). Agents adapt once they have had a chance to learn about the new rules, and may discover new methods to circumvent the rules. This paper presents a research strategy - composed of an experimental design and creative data collection - that permits an investigation of the effects of regulation net of agent adaptive behaviors.[17] This research approach should be broadly useful for policy evaluation whenever agents can react to circumvent enforcement. Further, we implement this research design to answer a policy question that is important in its own right: How best to curb illegal sales of a fish whose population is about to collapse? Such ecological threats have unfortunately become increasingly common around the world.

---

[17] An alternative evaluation strategy would be to collect data in the short run before agents have an opportunity to react to the new regime, and in the long-run after they have reacted. This is more expensive, requires more time, and fundamentally more difficult, because researchers do not always know when and how agents would learn and adapt.

Our experimental variations that change the specific attributes of enforcement policy yield novel empirical insights about the behavior of regulated agents, and how to better design policy accounting for the potential reactions of agents. Data collected via mystery shoppers help us identify the ways in which agents exploit loopholes to continue selling fish illegally. As a result, the standard measure of illegal activity (sales of fresh, visible hake) overstates the true effect of enforcement. Monitoring on an unpredictable schedule makes it more difficult for agents to circumvent enforcement. This, in turn, makes that strategy the most cost-effective way to reduce hake sales even though it is more expensive to implement. In contrast, a high frequency monitoring schedule produces a surprising result: it allows vendors to learn the regulators' strategies faster, and more effectively cheat, thereby undermining enforcement efforts.

We use multiple surveys of different market actors to document that these interventions travel downstream to affect consumer behavior and travel upstream to affect the behavior of fishermen who supply to vendors. They also travel laterally to affect vendors in control markets who are linked to treated markets either socially or spatially. Our investigation of vendor reactions through mystery shoppers, spillover effects on other market actors, and benchmarking these results against the effects of an information campaign, all combine to produce a comprehensive evaluation of an important environmental program. Ultimately we learn that enforcement can be the most cost-effective strategy for the Chilean government to pursue to curb illegal sales of hake fish, but that details of the enforcement policy design matter. Without sophisticated design-thinking, attempts at enforcement can backfire. Designing and implementing a consumer information campaign is a much less complex task, and many regulators may rationally choose to proceed with such simpler approaches.

# References

James Alm, Betty R Jackson, and Michael McKee. Getting the word out: Enforcement information dissemination and compliance behavior. *Journal of Public Economics*, 93(3):392–402, 2009.

Abhijit Banerjee, Esther Duflo, Raghabendra Chattopadhyay, Daniel Keniston, and Nina Singh. The efficient deployment of police resources: Theory and new evidence from a randomized drunk driving crackdown in india. 2017.

Gary Becker. Crime and punishment: An economic approach. *Journal of Political Economy*, 76: 169–217, 1968.

Christopher Blattman, Donald Green, Daniel Ortega, and Santiago Tobón. Pushing crime around the corner? estimating experimental impacts of large-scale security interventions. 2017.

Robin Burgess and Nicholas Stern. Taxation and development. *Journal of Economic Literature*, 31(2):762–830, 1993.

Paul Carrillo, Dina Pomeranz, and Monica Singhal. Dodging the taxman: Firm misreporting and limits to tax enforcement. *American Economic Journal: Applied Economics*, 9(2):144–164, 2017.

Raj Chetty, Mushfiq Mobarak, and Monica Singhal. Increasing tax compliance through social recognition. *Policy Brief*, 2014.

Esther Duflo, Michael Greenstone, Rohini Pande, and Nicholas Ryan. Truth-telling by third-party auditors and the response of polluting firms: Experimental evidence from india. *The Quarterly Journal of Economics*, 128(4):1499–1545, 2013.

Esther Duflo, Michael Greenstone, Rohini Pande, and Nicholas Ryan. The value of discretion in the enforcement of regulation: Experimental evidence and structural estimates from environmental inspections in india. *Econometrica (forthcoming)*, 2018.

Jan Eeckhout, Nicola Persico, and Petra E Todd. A theory of optimal random crackdowns. *American Economic Review*, 100(3):1104–35, 2010.

Haichao Fan, Yu Liu, Nancy Qian, and Jaya Wen. The dynamic effects of computerized vat invoices on chinese manufacturing firms. 2018.

FAO. Food and agriculture organization of the united nations, the state of world fisheries and aquaculture, 2014.

Raymond Fisman and Shang-Jin Wei. Tax rates and tax evasion: Evidence from "missing imports" in china. *Journal of Political Economy*, 112(2):471–496, 2004.

Edward L Glaeser and Andrei Shleifer. The rise of the regulatory state. *Journal of Economic Literature*, 41(2):401–425, 2003.

Wayne B Gray and Jay P Shimshack. The effectiveness of environmental monitoring and enforcement: A review of the empirical evidence. *Review of Environmental Economics and Policy*, 5(1): 3–24, 2011.

Raymond Guiteras, James Levinsohn, and Ahmed Mushfiq Mobarak. Encouraging sanitation investment in the developing world: A cluster-randomized trial. *Science*, 348(6237):903–906, 2015.

Benjamin Hansen. Punishment and deterrence: Evidence from drunk driving. *American Economic Review*, 105(4):1581–1617, 2015.

Winston Harrington. Enforcement leverage when penalties are restricted. *Journal of Public Economics*, 37(1):29–53, 1988.

Seema Jayachandran, Joost de Laat, Eric F Lambin, Charlotte Y Stanton, Robin Audy, and Nancy E Thomas. Cash for carbon: A randomized trial of payments for ecosystem services to reduce deforestation. *Science*, 357(6348):267–273, 2017.

Ginger Zhe Jin and Phillip Leslie. The effect of information on product quality: Evidence from restaurant hygiene grade cards. *The Quarterly Journal of Economics*, 118(2):409–451, 2003.

Anja Kollmuss and Julian Agyeman. Mind the gap: Why do people act environmentally and what are the barriers to pro-environmental behavior? *Environmental Education Research*, 8(3): 239–260, 2002.

Robert E Lucas. Econometric policy evaluation: A critique. In *Carnegie-Rochester conference series on public policy*, volume 1, pages 19–46. Elsevier, 1976.

Edward Miguel and Michael Kremer. Worms: Identifying impacts on education and health in the presence of treatment externalities. *Econometrica*, 72(1):159–217, 2004.

Deniz Okat. Deterring fraud by looking away. *The RAND Journal of Economics*, 47(3):734–747, 2016.

Benjamin A Olken. Monitoring corruption: Evidence from a field experiment in indonesia. *Journal of Political Economy*, 115(2):200–249, 2007.

Dina Pomeranz. No taxation without information: Deterrence and self-enforcement in the value added tax. *American Economic Review*, 105(8):2539–69, 2015.

Ritva Reinikka and Jakob Svensson. Fighting corruption to improve schooling: Evidence from a newspaper campaign in uganda. *Journal of the European Economic Association*, 3(2-3):259–267, 2005.

Abebe Shimeles, Daniel Zerfu Gurara, and Firew Woldeyes. Taxman's dilemma: Coercion or persuasion? evidence from a randomized field experiment in ethiopia. *American Economic Review*, 107(5):420–24, 2017.

Jay P Shimshack and Michael B Ward. Regulator reputation, enforcement, and environmental compliance. *Journal of Environmental Economics and Management*, 50(3):519–540, 2005.

Robert N. Stavins. The problem of the commons: Still unsettled after 100 years. *American Economic Review*, 101(1):81–108, 2011.

Subpesca. Propuesta de política pública de desarrollo productivo para la pesca artesanal. *Servicio Nacional de Pesca de Chile*, 2013.

Subpesca. Estado de situación principales pesquerías chilenas. *Servicio Nacional de Pesca de Chile*, 2015.

Robert M Townsend. Optimal contracts and competitive markets with costly state verification. *Journal of Economic Theory*, 21(2):265–293, 1979.

WWF. Estimación de la pesca inn en la pesquería de merluza común. *World Wide Fund for Nature*, 2017.

# Appendix

## A  Appendix Figures on the Research Context

### A.1  Fishermen Villages

Figure A.1: Fishermen Village (Caleta)

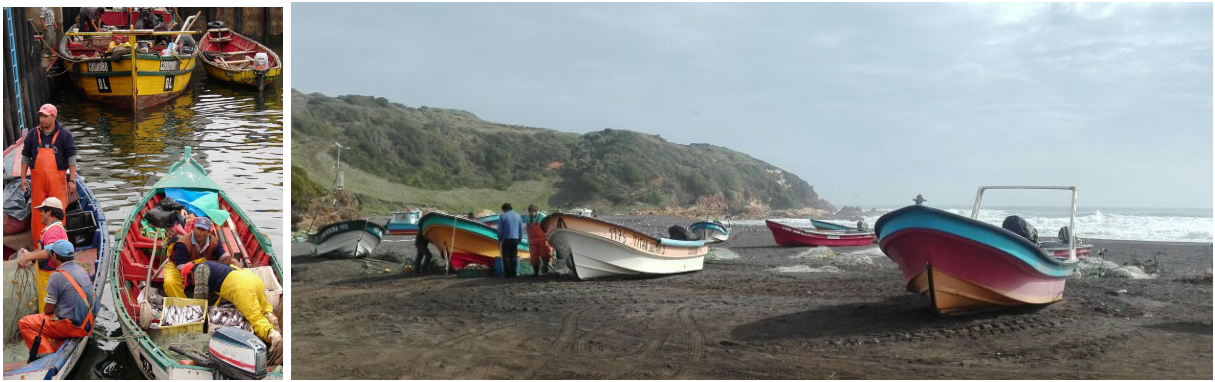

### A.2  Outdoor Markets

Figure A.2: Examples of Ferias

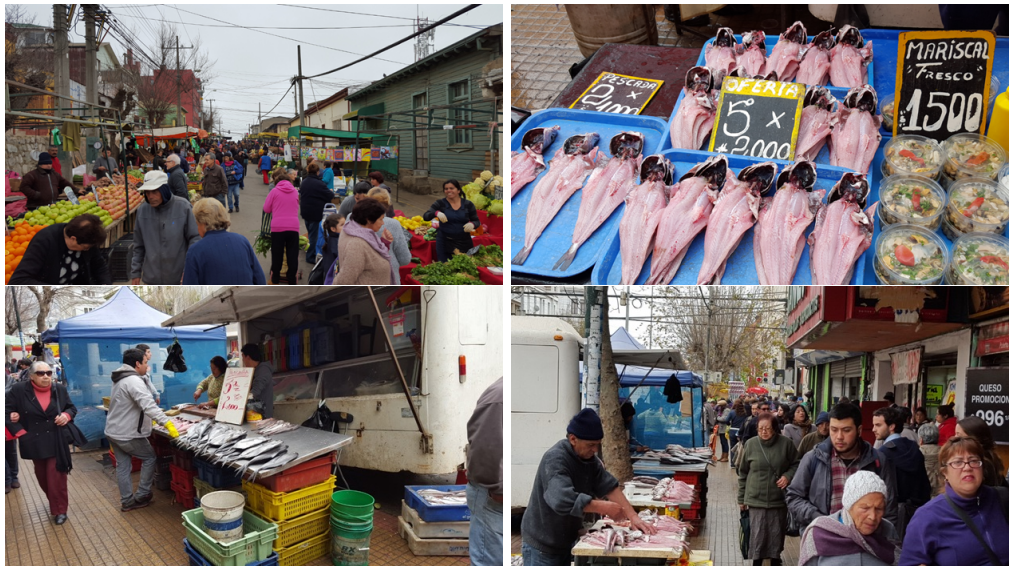

Figure A.3: Example of a Circuit

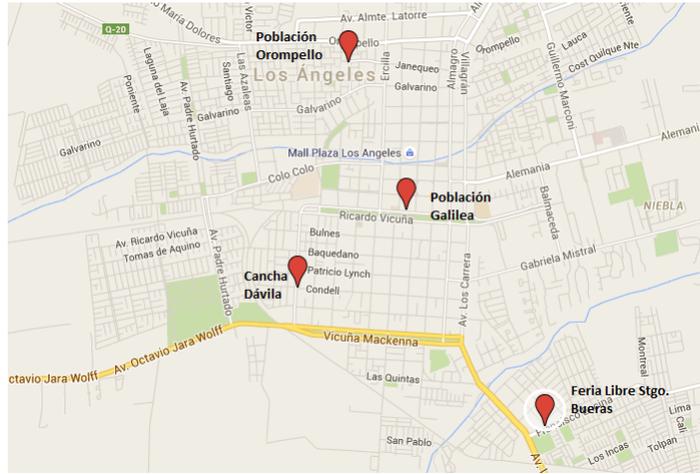

Figure A.3 maps the four ferias that compound one circuit of the city of Los Angeles, VII region.

Figure A.4: Map of Circuits and Caletas

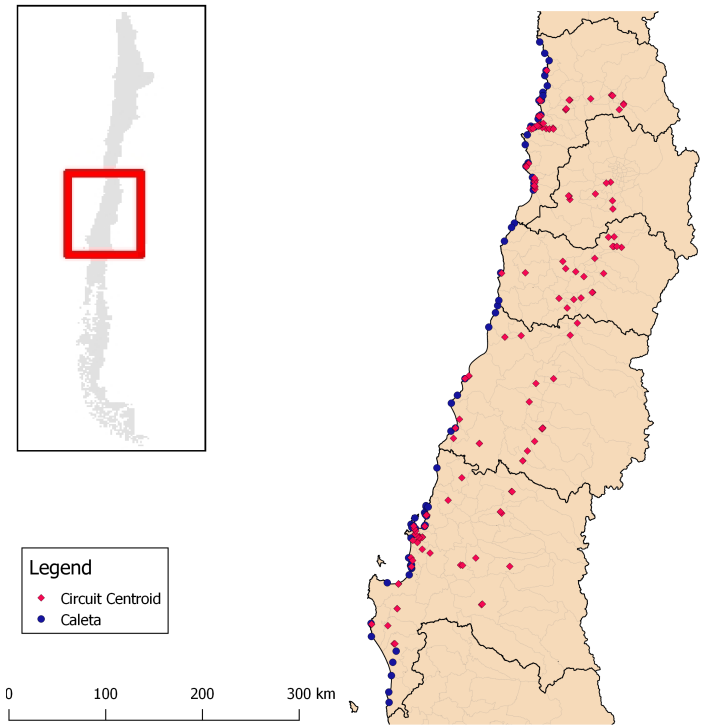

# B  Interventions

Figure B.1: Flyers

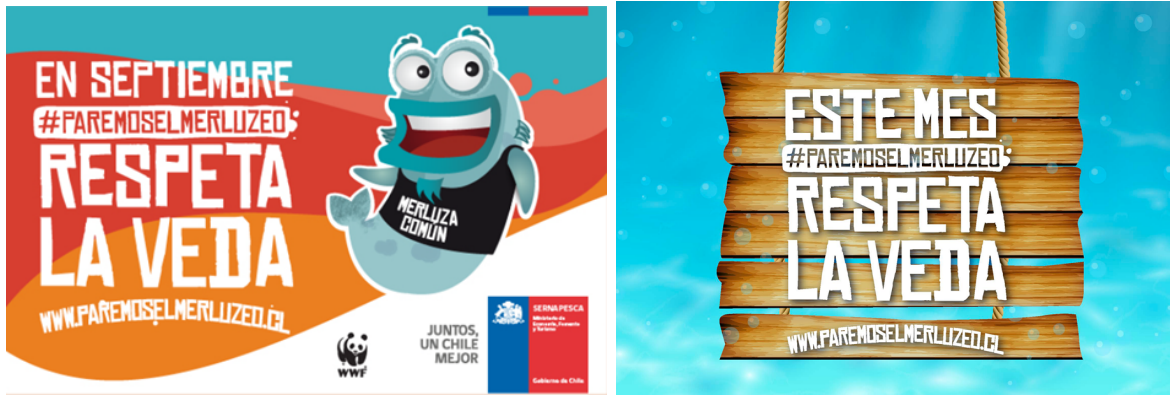

The figure B.1 shows the two types of flyers distributed during the ban period. The message of the one in the right says, "In September respect the Ban", the one in the right says "This month respect the Ban".

Figure B.2: Letter to Consumers

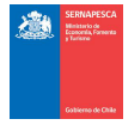

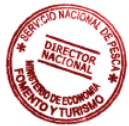

The figure B.2 shows the letter distributed to households during September 2015. The letter, signed by Sernapesca's director, informs about the September ban and the fact that hake's conservation is threatened because of overfishing.

Figure B.3: Examples of Neighborhood Treatment Assignment

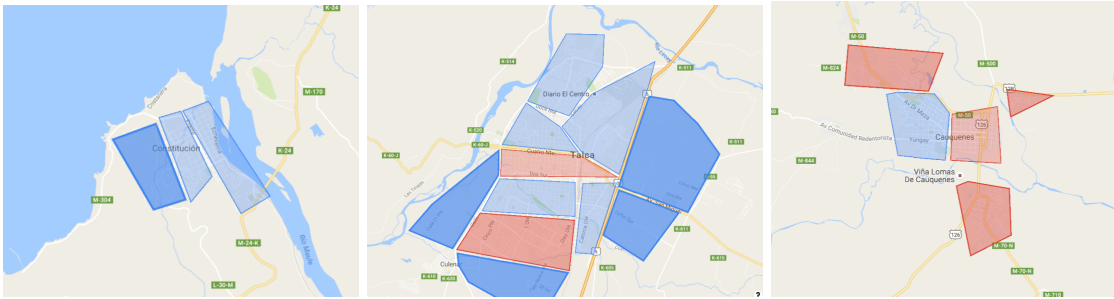

The 48 most populated comunas were divided randomly into three levels of saturation: high, low and zero. Based on the level of saturation, the information campaign was assigned at the neighborhood level. The figure B.3 shows the map of three different comunas: The comuna in the left didn't receive information campaign, the one in the center received low level of saturation, the one in the right received high level of saturation. In red, those neighborhoods assigned to receive the information campaign.

# C  Appendix Tables

## C.1  Descriptive Statistics

### C.1.1  Data Collected by Mystery Shoppers

During September 2015, the mystery shoppers interacted with fish-vendors 908 times. The table C.1 describes observable characteristics of the stalls visited and the vendors. In general, each stall was operated by one person. Mostly man, and based on mystery shoppers' guess, 47-year-old. The type of weight used informs about the formality of the stall; digital weights are more precise and expensive.

Table C.1: Fish Vendors in Ferias

| Variable | Mean | SD | Min | Max | N |
|---|---|---|---|---|---|
| Number of Vendors per Stall | 1.089 | 0.326 | 1 | 4 | 883 |
| Proportion Female Fish Vendor | 0.425 | 0.479 | 0 | 1 | 882 |
| Age Vendor | 47.438 | 10.126 | 19 | 75 | 876 |
| Prices Visibly Listed | 0.242 | 0.429 | 0 | 1 | 908 |
| Type of Weight | | | | | |
| *No Weight* | 0.262 | 0.440 | 0 | 1 | 848 |
| *Mechanical Weight* | 0.410 | 0.492 | 0 | 1 | 848 |
| *Digital Weight* | 0.308 | 0.462 | 0 | 1 | 848 |

This table presents observable characteristics of fish stalls visited by mystery shoppers during September 2015. The variable "Age Vendor" was not directly asked but guessed by mystery shoppers. The type of weight used to weight fish is a proxy of the level of formality of the fish stall.

The table C.2 describes the availability of different types of fish in feria stalls during the ban period. During a typical month, hake would be available in roughly 90% of stalls, however, due to the ban period (and our interventions), only 26% of stalls had hake for sale. The fish species offered in markets depend largely on the latitude of the market, i.e., markets located in the southern regions offer slightly different fish species than the stalls located in northern regions.

Table C.2: Fish Availability in Feria Stalls

| Fish | Availability | Price/unit (USD) | Price/kg (USD) | Unitary Weight (kg) | N |
|---|---|---|---|---|---|
| Hake | 0.263 | 1.08 | 3.81 | 0.284 | 239 |
| Pomfret | 0.684 | 5.27 | 4.99 | 1.057 | 621 |
| Mackerel | 0.124 | 2.05 | 4.16 | 0.492 | 113 |
| Silverside | 0.096 | 0.17 | 2.92 | 0.059 | 87 |
| Salmon | 0.139 | 7.53 | 9.22 | 0.816 | 126 |
| Sawfish | 0.057 | 6.27 | 6.25 | 1.003 | 52 |
| Albacore | 0.051 | . | 9.21 | . | 46 |
| Southern Hake | 0.042 | 7.30 | 5.59 | 1.306 | 38 |

This table presents the availability and average prices of different fish types in feria stalls during September 2015. The mystery shoppers recorded the price for each fish offered for sale in each fish-stall visited. The sale price in each stall was based on units, kilos or both. The unitary weight is estimated using the ratio of these two prices. The albacore is a considerably larger fish type (over 20 kgs) and is only sold in pieces (by kg).

### C.1.2 Data Collected in the Fisherman Survey

A round of surveys to Fishermen was collected in August 2016. In total, 231 fishermen were surveyed and asked about their work, typical buyers and fishing behavior. The table C.3 describes the main variables collected in the survey.

Table C.3: Fishermen Characteristics

| Variable | Mean | SD | Min | Max | N |
|---|---|---|---|---|---|
| *Fisherman Boat characteristics:* | | | | | |
|   Boat Length (mts) | 8.52 | 3.11 | 6 | 24 | 227 |
|   Boat Powered by a Motor | 0.88 | 0.33 | 0 | 1 | 231 |
|   Fiberglass Boat | 0.57 | 0.50 | 0 | 1 | 228 |
|   Wooden Boat | 0.39 | 0.49 | 0 | 1 | 228 |
| *Union Participation:* | | | | | |
|   Number of Unions in the Caleta | 1.67 | 0.90 | 0 | 3 | 230 |
|   Fisherman Member of a Union | 0.82 | 0.38 | 0 | 1 | 230 |
| *Number of Days that Goes Fishing Every Week:* | | | | | |
|   Summer | 5.01 | 1.47 | 1 | 7 | 226 |
|   Winter | 2.25 | 1.14 | 0 | 7 | 227 |
| *Number of Boats in the Caleta:* | | | | | |
|   Less than 10 | 0.24 | 0.43 | 0 | 1 | 189 |
|   Between 10 and 30 | 0.22 | 0.41 | 0 | 1 | 189 |
|   Between 31 and 60 | 0.24 | 0.43 | 0 | 1 | 189 |
|   Between 61 and 100 | 0.12 | 0.32 | 0 | 1 | 189 |
|   More than 100 | 0.19 | 0.39 | 0 | 1 | 189 |
| *Top 3 Most Captured Fishes in the Caleta:* | | | | | |
|   Hake | 0.56 | 0.50 | 0 | 1 | 230 |
|   Sawfish | 0.24 | 0.43 | 0 | 1 | 230 |
|   Cuttlefish | 0.24 | 0.43 | 0 | 1 | 230 |
|   Pomfret | 0.13 | 0.34 | 0 | 1 | 230 |
|   Bass | 0.10 | 0.30 | 0 | 1 | 230 |
| *Usual Buyer of the Fish at the Dock:* | | | | | |
|   Final Consumer | 0.58 | 0.49 | 0 | 1 | 230 |
|   Feria Vendor | 0.27 | 0.45 | 0 | 1 | 228 |
|   Intermediary | 0.60 | 0.49 | 0 | 1 | 227 |

This table describes the responses to the Fishermen Survey carried out in August 2016 to 231 fishermen. On average, three fishermen were surveyed in each of the 74 Caletas that operate in the four coastal regions included in our sample. The last section of the table represents the proportion that responded that *Always* or *Most of the Time* the fish was sold to these type of buyers.

## C.2 Balance Tables

The table C.4 presents the balance of relevant characteristics across different treatment arms. These variables include market's observable characteristics, socioeconomic characteristics of the municipality and weather information of the day of the visit by a mystery shopper. The columns 1, 2, 4 and 6 present the mean and SD of these variables for different treatment arms. The columns 3, 5 and 7 compare the difference relative to the control group as well as its p-value. Finally, joint significance tests are also reported in the last two columns. The tables C.5 and C.6 present the same estimates but decomposing by the enforcement variations: predictability and intensity.

Table C.4: Randomization Balance on Observables

|  | (1) | (2) | (3) | (4) | (5) | (6) | (7) |
|---|---|---|---|---|---|---|---|
| Variable | Control | Info Campaign | | Enforcement | | Enforc. and Info Camp. | |
|  | Mean | Mean | Diff | Mean | Diff | Mean | Diff |
| Indicator Fixed Stalls | 0.573 | 0.644 | 0.083 | 0.489 | -0.069 | 0.509 | 0.013 |
|  | ( 0.497) | ( 0.484) | [ 0.740] | ( 0.501) | [ 0.574] | ( 0.501) | [ 0.920] |
| Distance to Closest Caleta (kms) | 16.507 | 11.572 | 3.465 | 14.863 | -3.606 | 26.425 | 2.245 |
|  | ( 25.082) | ( 9.503) | [ 0.516] | ( 22.626) | [ 0.386] | ( 29.037) | [ 0.682] |
| Poverty Rate Municipality | 19.006 | 17.567 | -2.148 | 18.026 | -1.079 | 16.734 | -0.316 |
|  | ( 4.780) | ( 3.313) | [ 0.244] | ( 5.483) | [ 0.412] | ( 7.549) | [ 0.851] |
| Av. Monthly Income Municipality (USD) | 791.767 | 875.475 | 18.446 | 790.514 | -2.506 | 830.683 | 20.464 |
|  | ( 149.808) | ( 172.858) | [ 0.846] | ( 140.334) | [ 0.953] | ( 139.251) | [ 0.673] |
| Deliquency Rate Municipality | 0.038 | 0.029 | -0.013** | 0.036 | -0.001 | 0.034 | -0.004 |
|  | ( 0.015) | ( 0.002) | [ 0.016] | ( 0.015) | [ 0.835] | ( 0.009) | [ 0.480] |
| Rain Indicator | 0.290 | 0.133 | -0.124 | 0.178 | -0.114 | 0.142 | -0.122 |
|  | ( 0.456) | ( 0.344) | [ 0.455] | ( 0.383) | [ 0.318] | ( 0.349) | [ 0.301] |
| Average Temperature (Celsius) | 12.200 | 12.126 | 0.081 | 11.993 | -0.192 | 11.936 | -0.688 |
|  | ( 2.281) | ( 2.087) | [ 0.942] | ( 2.021) | [ 0.797] | ( 2.196) | [ 0.346] |
| Joint Significance | | | | | | | |
| F statistic | | | 0.609 | | 1.094 | | 0.816 |
| p-value | | | 0.747 | | 0.371 | | 0.575 |

This table reports characteristics of circuits included in our sample across treatment arms. The columns (1), (2), (4) and (6) show the mean and the standard deviation for the control and treatment groups. The columns (3), (5) and (7) show the coefficient on treatments from regressions of each characteristic on treatments and strata fixed effects, clustreing standard errors at the circuit level. The p-values are reported in brackets. The socio-economic characteristics are aggregated at Municipality level. These variables should be interpreted as the characteristics of the Municipality where the circuit is located. Also, this table reports weather information of the day that different circuits were visited by mystery shoppers. Finally, joint significance test statistics: F statistic and p-values, for all variables on each treatment arm are reported in the last two rows of the table. *** p<0.01, ** p<0.05, * p<0.1

Table C.5: Randomization Balance: Enforcement Predictability

| Variable | (1) Control Mean | (2) Info Campaign Mean | (3) Info Campaign Diff | (4) Enforc: Predictable Schedule Mean | (5) Enforc: Predictable Schedule Diff | (6) Enforc: Unpredictable Schedule Mean | (7) Enforc: Unpredictable Schedule Diff |
|---|---|---|---|---|---|---|---|
| Indicator Fixed Stalls | 0.573 | 0.644 | 0.080 | 0.396 | -0.148 | 0.575 | 0.045 |
|  | ( 0.497) | ( 0.484) | [ 0.749] | ( 0.490) | [ 0.270] | ( 0.495) | [ 0.710] |
| Distance to Closest Caleta (kms) | 16.507 | 11.572 | 3.822 | 17.286 | 1.704 | 20.487 | -4.926 |
|  | ( 25.082) | ( 9.503) | [ 0.479] | ( 23.417) | [ 0.678] | ( 27.384) | [ 0.333] |
| Poverty Rate Municipality | 19.006 | 17.567 | -2.173 | 17.130 | -1.739 | 17.890 | -0.075 |
|  | ( 4.780) | ( 3.313) | [ 0.240] | ( 6.185) | [ 0.212] | ( 6.450) | [ 0.960] |
| Av. Monthly Income Municipality (USD) | 791.767 | 875.475 | 18.828 | 809.457 | -1.068 | 801.805 | 9.235 |
|  | ( 149.808) | ( 172.858) | [ 0.843] | ( 153.747) | [ 0.981] | ( 130.506) | [ 0.834] |
| Deliquency Rate Municipality | 0.038 | 0.029 | -0.013** | 0.036 | -0.002 | 0.035 | -0.001 |
|  | ( 0.015) | ( 0.002) | [ 0.015] | ( 0.015) | [ 0.648] | ( 0.012) | [ 0.782] |
| Rain Indicator | 0.290 | 0.133 | -0.126 | 0.117 | -0.158 | 0.202 | -0.080 |
|  | ( 0.456) | ( 0.344) | [ 0.446] | ( 0.322) | [ 0.162] | ( 0.402) | [ 0.488] |
| Average Temperature (Celsius) | 12.200 | 12.126 | 0.076 | 12.058 | -0.170 | 11.904 | -0.491 |
|  | ( 2.281) | ( 2.087) | [ 0.946] | ( 2.057) | [ 0.824] | ( 2.107) | [ 0.498] |
| Joint Significance |  |  |  |  |  |  |  |
| $F$ statistic |  |  | 0.609 |  | 1.954 |  | 1.717 |
| p-value |  |  | 0.747 |  | 0.067 |  | 0.111 |

This table reports characteristics of circuits included in our sample across treatment arms. The columns (1), (2), (4) and (6) show the mean and the standard deviation for the control and treatment groups. The columns (3), (5) and (7) show the coefficient on treatments from regressions of each characteristic on treatments and strata fixed effects, clustering standard errors at the circuit level. The p-values are reported in brackets. The socio-economic characteristics are aggregated at Municipality level. These variables should be interpreted as the characteristics of the Municipality where the circuit is located. Also, this table reports weather information of the day that different circuits were visited by mystery shoppers. Finally, joint significance test statistics: F statistic and p-values, for all variables on each treatment arm are reported in the last two rows of the table. *** p<0.01, ** p<0.05, * p<0.1

Table C.6: Randomization Balance: Enforcement Intensity

|  | (1) | (2) | (3) | (4) | (5) | (6) | (7) |
|---|---|---|---|---|---|---|---|
| Variable | Control | Info Campaign | | Enforc: High Intensity | | Enforc: Low Intensity | |
|  | Mean | Mean | Diff | Mean | Diff | Mean | Diff |
| Indicator Fixed Stalls | 0.573 | 0.644 | 0.098 | 0.372 | -0.164 | 0.591 | 0.039 |
|  | ( 0.497) | ( 0.484) | [ 0.696] | ( 0.484) | [ 0.211] | ( 0.492) | [ 0.753] |
| Distance to Closest Caleta (kms) | 16.507 | 11.572 | 4.009 | 16.224 | -5.302 | 21.269 | 0.600 |
|  | ( 25.082) | ( 9.503) | [ 0.451] | ( 23.309) | [ 0.330] | ( 27.292) | [ 0.885] |
| Poverty Rate Municipality | 19.006 | 17.567 | -1.975 | 16.242 | -2.256 | 18.564 | 0.140 |
|  | ( 4.780) | ( 3.313) | [ 0.279] | ( 7.011) | [ 0.101] | ( 5.576) | [ 0.922] |
| Av. Monthly Income Municipality (USD) | 791.767 | 875.475 | 17.128 | 821.315 | 23.237 | 792.779 | -8.734 |
|  | ( 149.808) | ( 172.858) | [ 0.857] | ( 142.660) | [ 0.587] | ( 138.930) | [ 0.840] |
| Deliquency Rate Municipality | 0.038 | 0.029 | -0.013** | 0.036 | -0.000 | 0.035 | -0.003 |
|  | ( 0.015) | ( 0.002) | [ 0.014] | ( 0.013) | [ 0.973] | ( 0.014) | [ 0.547] |
| Rain Indicator | 0.290 | 0.133 | -0.128 | 0.193 | -0.077 | 0.143 | -0.143 |
|  | ( 0.456) | ( 0.344) | [ 0.439] | ( 0.396) | [ 0.491] | ( 0.350) | [ 0.215] |
| Average Temperature (Celsius) | 12.200 | 12.126 | 0.050 | 12.127 | -0.192 | 11.853 | -0.447 |
|  | ( 2.281) | ( 2.087) | [ 0.964] | ( 2.204) | [ 0.799] | ( 1.984) | [ 0.545] |
| Joint Significance |  |  |  |  |  |  |  |
| *F statistic* |  |  | 0.609 |  | 2.016 |  | 2.090 |
| *p-value* |  |  | 0.747 |  | 0.058 |  | 0.049 |

This table reports characteristics of circuits included in our sample across treatment arms. The columns (1), (2), (4) and (6) show the mean and the standard deviation for the control and treatment groups. The columns (3), (5) and (7) show the coefficient on treatments from regressions of each characteristic on treatments and strata fixed effects, clustering standard errors at the circuit level. The p-values are reported in brackets. The socio-economic characteristics are aggregated at Municipality level. These variables should be interpreted as the characteristics of the Municipality where the circuit is located. Also, this table reports weather information of the day that different circuits were visited by mystery shoppers. Finally, joint significance test statistics: F statistic and p-values, for all variables on each treatment arm are reported in the last two rows of the table. *** p<0.01, ** p<0.05, * p<0.1

## C.3  Exit of Stalls Correction

Our main results are based on the information gathered by mystery shoppers from the operative stalls at the moment of the visit, which does not capture the fact that the "missing" stalls are not selling hake. To correct for this issue we identify the average number of stalls per circuit/visit before and after the interventions. The comparison between these two averages informs about the number of "missing" stalls per circuit.[18] The number of stalls observed by mystery shoppers in every visit in the post treatment period is increased by computed number of missing stalls. The

---
[18] We allow the number of missing stalls to be non-integer, and negative if the number of stalls increased.

added observations have zero fish available.[19] [20]

Table C.7: Treatment Effect on Hake Availability Correcting for the Exit of Stalls

|  | (1) Fresh, Visible Hake | (2) Any Hake Available (Fresh-Visible, Hidden or Frozen) |
|---|---|---|
| **Panel A: Main Specification** | | |
| Info Campaign Only | -0.118** | -0.115* |
|  | (0.060) | (0.065) |
| Enforcement Only | -0.190** | -0.141 |
|  | (0.082) | (0.091) |
| Info Campaign and Enforcement | -0.156* | -0.130 |
|  | (0.084) | (0.104) |
| **Panel B: Variation in Predictability of Enforcement** | | |
| Info Campaign Only | -0.111* | -0.121* |
|  | (0.062) | (0.064) |
| Enforcement on Predictable Schedule | -0.091 | -0.061 |
|  | (0.073) | (0.087) |
| Enforcement on Unpredictable Schedule | -0.246*** | -0.197** |
|  | (0.089) | (0.100) |
| **Panel C: Variation in Frequency of Enforcement** | | |
| Info Campaign Only | -0.113* | -0.121* |
|  | (0.062) | (0.064) |
| High Intensity Enforcement | -0.086 | -0.092 |
|  | (0.092) | (0.101) |
| Low Intensity Enforcement | -0.184** | -0.148 |
|  | (0.086) | (0.095) |
|  | | |
| Change in Dep Var in Control | | |
| During Intervention | -0.17 | -0.28 |
| Covariates | Yes | Yes |
| Baseline Control | Yes | Yes |
| N | 1014 | 1014 |

This table presents the coefficient corresponding to the interaction term $T_c \times Post_t$ for each treatment correcting for the exit of stalls. The increase in the number of observations relative to results presented earlier is due to the fact that the correction is done by adding the "missing" stalls (calculated comparing the number of stalls per circuit before and after the interventions). The panel A describes the same specification presented in Table 3. Panels B and C follow the same specification than Table 5. Probit regression marginal effects are reported. Robust standard errors clustered by circuit in parentheses.*** p<0.01, ** p<0.05, * p<0.1

---

[19] If the number "missing" stalls is negative: the number of stalls observed by mystery shoppers in every visit in the pre-treatment period is increased by that number.

[20] Since we allow the number "missing" stalls to be non-integer, we add a random noise that distributes uniform between -0.5 and 0.5, and then, the sum of the "missing" number and the noise is rounded to the closest integer. This correction makes the expansion more representative of the right (possibly non-integer) number.

## C.4 Longer-Run Effects

Table C.8: Longer term Effects: Hake Purchases Reported by Consumers in March 2016 (Outside Ban Period)

| VARIABLES | (1) Purchased Hake last month | (2) Number Times Hake Purchased | (3) Usually Purchase Hake |
|---|---|---|---|
| Information Campaign Only | -0.127 | -0.420* | -0.116 |
|  | (0.100) | (0.235) | (0.095) |
| Enforcement Only | 0.017 | -0.107 | 0.090 |
|  | (0.075) | (0.206) | (0.067) |
| Info Campaign and Enforcement | -0.016 | -0.198 | 0.017 |
|  | (0.077) | (0.250) | (0.074) |
| Mean Dep Var Control | 0.52 | 0.98 | 0.51 |
| N | 3652 | 3630 | 3652 |

This table presents the effect of different treatment arms on the reported consumption of hake fish by consumers based on the round of surveys collected in March 2016. Columns 1 and 3 show marginal effects from Probit regressions. Column 2 shows the marginal effects from a Poisson regression because the dependent variable is a count data. These regressions control for propensity to purchase other types of fishes and other covariates. Standard errors are clustered based on the circuit where the survey was collected. *** p<0.01, ** p<0.05, * p<0.1

Table C.9: Longer Term Effects: Hake Availability in Ferias in March 2016 (Outside Ban Period)

| VARIABLES | (1) Fresh, Visible Hake | (2) Any Hake Available (Fresh-Visible, Hidden or Frozen) |
|---|---|---|
| Information Campaign Only | 0.000 | -0.009 |
|  | (0.104) | (0.085) |
| Enforcement Only | -0.095 | -0.077 |
|  | (0.063) | (0.066) |
| Info Campaign and Enforcement | -0.084 | -0.099 |
|  | (0.080) | (0.085) |
| Mean Dep Var Control | 0.86 | 0.90 |
| N | 754 | 754 |

This table reports the effect of each treatment arm on the availability of hake six months after the interventions. Each observation correspond to a single stall visited by a mystery shopper. Circuits were visited twice during March 2016. The table reports marginal effects from Probit regression including covariates. Robust standard errors are clustered by circuit, which was the unit of randomization. *** $p<0.01$, ** $p<0.05$, * $p<0.1$

## C.5 Consumer Mobility Between Neighborhoods

The table C.10 shows the proportion of consumers treated by the information campaign depending on the location of the feria where they are surveyed. The striking fact in this table is that in high-saturation municipalities, the proportion of consumers treated with the information campaign is high, regardless of whether we found that person shopping in a feria located in a treatment neighborhood (78%) or in a control neighborhood (69%). High Information campaign saturation is therefore the effective treatment variable, and conditional on that, the specific location of the feria does not matter too much.

Table C.10: Proportion of Consumers located in Treated Neighborhoods

|  | Survey in Feria located in Treated Neigh | | Survey in Feria located in Control Neigh | |
| --- | --- | --- | --- | --- |
|  | Prop | N | Prop | N |
| High Saturation Municipality | 0.78 | 1114 | 0.69 | 389 |
| Low Saturation Municipality | 0.57 | 559 | 0.17 | 825 |
| Zero Saturation Municipality | 0 | 0 | 0.00 | 1014 |
| Overall | 0.71 | 1673 | 0.18 | 2228 |

This table shows the proportion of consumers whose households are located in neighborhoods assigned to receive the information campaign. These statistics are based on households' location reported by surveyed consumers. 71% of consumers surveyed in a feria located in a treated neighborhood live in a household located in a treated neighborhood; the remaining 29% are consumers who live in a household located in a control neighborhood. This table informs about the high consumers' mobility between neighborhoods. In fact, in high-saturation municipalities, the proportion of treated consumers is higher in both, ferias located in treatment and control neighborhoods.

## C.6 Alternative Definition Information Campaign Treatment

Tables C.11 and C.12 present the main results using a different definition of the Information Campaign treatment: The variable "Information campaign" indicates whether the observations were collected by mystery shoppers in ferias located in treated neighborhoods - regardless of the level of saturation of the municipality. This definition does not include possible information spill-overs between neighborhoods within municipalities assigned to receive information.

Table C.11: Treatment Effect on Hake Availability

| VARIABLES | (1) Fresh, Visible Hake | (2) Any Hake Available (Fresh-Visible, Hidden or Frozen) |
|---|---|---|
| Information Campaign Only | -0.082 | -0.070 |
|  | ( 0.064) | ( 0.071) |
| Enforcement Only | -0.157** | -0.101 |
|  | ( 0.079) | ( 0.094) |
| Info Campaign and Enforcement | -0.169** | -0.121 |
|  | ( 0.079) | ( 0.094) |
|  |  |  |
| Change in Dep Var in Control |  |  |
| Markets During Intervention | -0.21 | -0.36 |
| N | 901 | 901 |

This table reports the effect of each treatment arm on the availability of illegal hake fish. The variable "Info Campaign" indicates if the feria where the data was collected is located in a neighborhood assigned to receive the information campaign. Probit Marginal effects of he interactions $T_c \times Post_t$ are reported. Robust standard errors are clustered by circuit and presented in parentheses. *** p<0.01, ** p<0.05, * p<0.1

Table C.12: Treatment Effect on Hake Sales by Enforcement Strategy

|  | (1) | (2) |
|---|---|---|
| VARIABLES | Fresh Hake | Any Hake |
| Panel A: Variation in Predictability of Enforcement Schedule | | |
| Info Campaign Only | -0.082 | -0.073 |
|  | ( 0.063) | ( 0.071) |
| Enforcement on Predictable Schedule | -0.075 | -0.036 |
|  | ( 0.071) | ( 0.089) |
| Enforcement on Unpredictable Schedule | -0.223** | -0.169* |
|  | ( 0.088) | ( 0.099) |
| Panel B: Variation in Frequency of Enforcement Visits | | |
| Info Campaign Only | -0.082 | -0.073 |
|  | ( 0.063) | ( 0.071) |
| High Intensity Enforcement | -0.045 | -0.049 |
|  | ( 0.087) | ( 0.101) |
| Low Intensity Enforcement | -0.177** | -0.140 |
|  | ( 0.086) | ( 0.095) |
| Change in Dep Var in control Markets | | |
| During Intervention | -0.21 | -0.36 |
| N | 901 | 901 |

This table reports the effect of each treatment arm on the availability of illegal hake fish. Panel A includes a dummy for the intensity sub-treatment, and Panel B includes a dummy for the predictability sub-treatment, but those coefficients are not shown. Each regression controls for the dependent variable in pre-intervention period, strata fixed effects and municipality characteristics. The variable "Info Campaign" indicates if the feria where the data was collected is located in a neighborhood assigned to receive the information campaign. Probit Marginal effects of he interactions $T_c \times Post_t$ are reported. Robust standard errors are clustered by circuit and presented in parentheses. *** p<0.01, ** p<0.05, * p<0.1

# D  Further Details on the Cost-Effectiveness Analysis

The section 7 describes the cost-effectiveness analysis of each treatment. These calculations are based on the following parameters:

- Costs: The total cost of implementing enforcement was $ 62,900.25, which is divided into fixed costs $ 7,338.06, and variable costs: $ 55,562.19. The fixed costs include administrative

staff salaries, central office coordination, etc. The variable costs include the specific costs incurred to implement the enforcement (i.e., financial compensation of inspectors, gasoline, etc.). Based on Sernapesca information, deploying enforcement in an unpredictable way is 10% more costly regarding staff availability. The cost of implementing enforcement in low intensity is obtained by calculating (variable) cost of each visit and multiplying by the number of visits under this new regime, adding the fixed costs.

The total cost of implementing the information campaign was $ 16,213.53, which includes the printing and distribution of flyers, posters, and letters in treated neighborhoods.

- Reduction of fish sales: The estimated effects of selling hake during the ban presented in section 5 are translated into numbers of fishes "saved." This exercise takes into account that every stall has 25 hake fishes available, there are 2.57 fish stalls in each feria. Each circuit operates 5 days a week, and the effects consider the three last weeks of September. The enforcement treatment contemplated 83 circuits, whereas 26 circuits are located in municipalities assigned to receive information campaign with a high level of saturation.